%% file: main.tex
\documentclass[11pt]{article}
\usepackage{amssymb,amsmath}
\usepackage{graphicx}
\usepackage{graphics}
\usepackage{subfigure} 
\usepackage{paralist}
\usepackage[T1]{fontenc}
\usepackage[latin1]{inputenc}    
\usepackage{pifont}    
\usepackage{eepic,epic}
\usepackage{epsfig}

 \makeatletter%
 \newcommand{\doublespacing}{\let\CS=
 \@currsize\renewcommand{\baselinestretch}{1.75}\tiny\CS}
 \newcommand{\extradoublespacing}{\let\CS=
 \@currsize\renewcommand{\baselinestretch}{1.9}\tiny\CS}
 \newcommand{\draftspacing}{\let\CS=
 \@currsize\renewcommand{\baselinestretch}{2.0}\tiny\CS}
 \newcommand{\hugedraftspacing}{\let\CS=
 \@currsize\renewcommand{\baselinestretch}{2.4}\tiny\CS}
 \makeatother%}
\newcommand{\OMIT}[1]{} % Suppress printing some part in your text.
\newcommand\qedblob{\ding{113}}
\def\literalqed{{\ \nolinebreak\hfill\mbox{\qedblob\quad}}}

\newenvironment{proofs}{\noindent{\bf Proof.}\hspace*{1em}}{\literalqed\bigskip}

\newcommand{\acc}[1]{\mathit{acc}_{#1}}
\newcommand{\solutiontrp}[1]{\mbox{\sc Sol}_{\scriptsize \trp}({#1})}

\newcommand{\shape}{\mathit{shape}}

\newcommand{\littlep}{{p}}
\newcommand{\parsimonious}{\ensuremath{\leq_{\mathit{par}}^{{\littlep}}}}
\newcommand{\randomized}{\ensuremath{\leq_{\mathit{ran}}^{{\littlep}}}}

\newcommand{\probbf}{\rm}
\newcommand{\sat}{{\probbf \mbox{SAT}}}
\newcommand{\usat}{{\probbf \mbox{Unique}\hbox{-}\allowbreak\mbox{SAT}}}
\newcommand{\circuitsat}{{\probbf \mbox{Circuit}\hbox{-}\allowbreak\mbox{SAT}}}
\newcommand{\circuitandnotsat}{{\probbf \mbox{Circuit}_{\wedge,\neg}\hbox{-}\allowbreak\mbox{SAT}}}
\newcommand{\trp}{{\probbf \mbox{TRP}}}
\newcommand{\utrp}{{\probbf \mbox{Unique}\hbox{-}\allowbreak\mbox{TRP}}}
\newcommand{\sharptrp}{{\probbf \#\mbox{TRP}}}

\newcommand{\np}{\mbox{\rm NP}}
\newcommand{\conp}{\mbox{\rm coNP}}
\newcommand{\DP}{\mbox{\rm DP}}
\newcommand{\sigmastar}{\mbox{$\Sigma^\ast$}}
\newcommand{\naturalnumber}{\ensuremath{{  \mathbb{N} }}}
\def\nats{\naturalnumber}

\newcommand{\condition}{\,|\:}

\newcommand{\sharpp}{{\rm \#P}}
\newcommand{\sharpsat}{{\probbf \#\mbox{SAT}}}

\newenvironment{desctight}
  {\begin{list}{}{\setlength\labelwidth{0pt}%
        \setlength{\itemsep}{0pt}%
        \setlength{\parsep}{0pt}%
        \setlength\itemindent{-\leftmargin}%
        }}
    {\end{list}}

  \newtheorem{theorem}{Theorem}[section]

  \newtheorem{lemma}[theorem]{Lemma}

  \newtheorem{definition}[theorem]{Definition}

\usepackage{ulem}
\begin{document}

\title{Satisfiability Parsimoniously Reduces to the 
Tantrix\texttrademark\ Rotation Puzzle
Problem\thanks{Supported in part by DFG grants RO~\mbox{1202/9-3} and
RO~\mbox{1202/11-1} and by
the Alexander von Humboldt Foundation's TransCoop program.
A preliminary version of this paper appeared in the Proceedings of
\emph{Machines, Computations and Universality} (MCU~2007).  URLs:
\mbox{\tt{}ccc.cs.uni-duesseldorf.de/\mbox{\tiny$\sim\,$}\{baumeister,
rothe\}} (D. Baumeister and J. Rothe).  
%Corresponding author:
%J. Rothe.
}  }

%\address{J\"{o}rg Rothe, Institut f\"ur Informatik,
% Heinrich-Heine-Universit\"at D\"usseldorf,
% 40225 D\"usseldorf, Germany.  E-mail: rothe@cs.uni-duesseldorf.de.}

\author{Dorothea Baumeister \textnormal{and} J\"{o}rg Rothe
\\ Institut f\"ur Informatik
\\ Heinrich-Heine-Universit\"at D\"usseldorf
\\ 40225 D\"usseldorf, Germany
}

\date{June 9, 2008}
%\setcounter{page}{1}
%\issue{123}

\maketitle
%\runninghead{D. Baumeister, J. Rothe}{Satisfiability Parsimoniously Reduces to the Tantrix\texttrademark\ Rotation Puzzle Problem}

\begin{abstract}
Holzer and Holzer~\cite{hol-hol:j:tantrix} proved that the
Tantrix\texttrademark\ rotation puzzle problem is $\np$-complete.
They also showed that for infinite rotation puzzles, this problem
becomes undecidable.  We study the counting version and the unique
version of this problem. 
% In particular, 
We prove that the
% (counting version of the) 
satisfiability problem parsimoniously
reduces to the
% (counting version of the) 
Tantrix\texttrademark\ rotation puzzle problem.
In particular, this reduction preserves the uniqueness of the
solution, which implies that the unique Tantrix\texttrademark\ 
rotation puzzle problem is as hard as the unique satisfiability problem,
and so is $\DP$-complete under polynomial-time randomized reductions,
where $\DP$ is the second level of the boolean hierarchy over~$\np$.
\\[2mm]
\end{abstract}

\section{Introduction}

Tantrix\texttrademark\ is a puzzle game played with hexagonal tiles
firmly arranged in the plane that each can be rotated around their
axes.  There are four different types of tiles (called \emph{Sint},
\emph{Brid}, \emph{Chin}, and \emph{Rond}, see
Figure~\ref{fig:TantrixTiles}) that differ by the form of the three
colored lines they each have, where the colors are chosen among
\emph{red}, \emph{yellow}, \emph{blue}, and \emph{green}.  The
objective of the game is to find a rotation of the given tiles so as
to create long lines and loops of the same color.  Since its invention
in 1991 by Mike McManaway from New Zealand and its commercial launch,
the Tantrix\texttrademark\ rotation puzzle has become extremely
popular and commercially successful.

Holzer and Holzer~\cite{hol-hol:j:tantrix} considered two variants of
the Tantrix\texttrademark\ rotation puzzle problem, one with finitely
many and one with infinitely many tiles in a given problem instance.
They proved that the finite variant of this problem is $\np$-complete
by reducing the $\np$-complete boolean circuit satisfiability problem
(restricted to circuits with AND and NOT gates only) to it.  They also
showed that the infinite variant of the Tantrix\texttrademark\
rotation puzzle problem is undecidable, again employing a circuit
construction.  For other results on the complexity of problems related
to Domino-like strategy games, we refer to
Gr{\"{a}}del~\cite{gra:j:domino}.

We consider two variants of the finite Tantrix\texttrademark\ rotation
puzzle problem, its counting version and its unique version.  The
% corresponding 
counting problem asks for the number of solutions of a
given rotation puzzle instance.  The
%  corresponding 
unique problem asks
whether a given rotation puzzle instance has exactly one solution.
Our main result is that the satisfiability problem parsimoniously
reduces to the Tantrix\texttrademark\ rotation puzzle problem.

The class $\sharpp$
% , which contains all functions giving the number of
% accepting computation paths of nondeterministic polynomial-time Turing
% machines,
was introduced by Valiant~\cite{val:j:permanent} to
capture the complexity of counting the solutions of $\np$
problems.  Parsimonious reductions between $\np$ counting
problems---such as ours---preserve the precise number of solutions.
This is an important property for at least two reasons.  First, the
structure of the solution space is preserved by a parsimonious
reduction from $A$ to~$B$, since solutions of $A$ are mapped
bijectively to solutions of $B$ in polynomial time.  Second,
parsimonious reductions can be used to prove lower bounds for the
unique versions of $\np$ problems.  In particular, we apply our 
above-mentioned parsimonious reduction to 
prove that the unique Tantrix\texttrademark\ rotation puzzle problem
is $\DP$-complete
under polynomial-time randomized reductions in the sense of Valiant
and Vazirani~\cite{val-vaz:j:np-unique}.
Here, $\DP$ is the set of differences of any two
$\np$ sets~\cite{pap-yan:j:dp}; so $\np \subseteq \DP$, and it is
considered most unlikely that both classes are equal.
Note that $\DP$ is the second level of the boolean hierarchy over~$\np$,
see Cai et al.\
\cite{cai-gun-har-hem-sew-wag-wec:j:bh1,cai-gun-har-hem-sew-wag-wec:j:bh2}.
Further results on $\DP$ and completenes in the  boolean hierarchy
over $\np$ can be found, e.g.,
in~\cite{cai-mey:j:dp,rot:j:exact-four-colorability},
see also the survey~\cite{rie-rot:j:boolean-hierarchies-survey}.

While many standard reductions between $\np$-complete problems are
easily seen to be parsimonious, there are a number of exceptions.  For
example, Barbanchon~\cite{bar:j:unique-graph-color} showed that the
(planar) satisfiability problem is parsimoniously polynomial-time
reducible to the (planar) $3$-colorability problem via a rather
sophisticated construction.  
Other examples of nontrivial parsimonious reductions can be found
in~\cite{pap:b-1994:complexity}.
Holzer and Holzer's reduction, however, is not
parsimonious~\cite{hol-hol:j:tantrix}.  The main purpose of this paper
is to show how to modify their reduction so as to make it
parsimonious.

We mention in passing that this paper differs from its preliminary
version~\cite{bau-rot:c:tantrix} in various ways.  First, to allow
comparison, we here explicitly show the differences between Holzer and
Holzer's original construction~\cite{hol-hol:j:tantrix} and our
modified construction by (a)~presenting their subpuzzles (marked so as
to clearly indicate the tiles that require modification if one aims at
a parsimonious reduction), and (b)~highlighting all modified or
additionally inserted tiles in our subpuzzles.  Second, unlike the
reductions given in~\cite{hol-hol:j:tantrix,bau-rot:c:tantrix}, in
the reduction presented here we
introduce a new subpuzzle for simulating wire crossings in
boolean circuits.  This new subpuzzle, which we call CROSS, will save
us the effort of transforming general boolean circuits into planar
boolean circuits (i.e., into circuits without wire crossings).  Hence, the
reduction provided in the present papers is more efficient and the
total number of tiles needed to simulate a given circuit
is considerably smaller than in our
previous construction~\cite{bau-rot:c:tantrix}.  Finally, to prove
correctness of our reduction, we now---unlike the approach taken 
in~\cite{bau-rot:c:tantrix}---argue via ``color sequences'' of
tiles, which facilitates reading and understanding the
arguments.

This paper is organized as follows.  In Section~\ref{sec:prelims}, we
define some complexity-theoretic notions and our variants of the
Tantrix\texttrademark\ rotation puzzle problem.  In
Section~\ref{sec:sharptrp-is-sharpp-complete}, we present our
parsimonious reduction.  In Section~\ref{sec:uniquetrp-is-dp-complete},
finally, we study the unique
version of this problem and show its $\DP$-completeness under
randomized reductions.

\section{Preliminaries}
\label{sec:prelims}

\subsection{Definition of Some Complexity-Theoretic Notions}

Fix the alphabet $\Sigma = \{0,1\}$, and let $\sigmastar$ denote the
set of strings over~$\Sigma$.  As is common, decision problems are
suitably encoded as languages over~$\Sigma$.  For any language $A
\subseteq \sigmastar$, let $\|A\|$ denote the number of elements
in~$A$.  For some background on computational complexity theory,
% the definition of well-known complexity classes and
% complexity-theoretic notions, 
we refer to any standard textbook of this field, e.g.,
\cite{pap:b-1994:complexity,rot:b:cryptocomplexity}.  Let $\np$ denote
the class of problems solvable in nondeterministic polynomial time.
Generalizing~$\np$, Papadimitriou and Yannakakis~\cite{pap-yan:j:dp}
introduced the class $\DP = \{A-B \condition A, B \in \np\}$ to
capture the complexity of $\np$-hard or $\conp$-hard problems that
seemingly are neither in $\np$ nor in~$\conp$.  In particular, they
showed that $\DP$ contains a number of \emph{uniqueness problems},
\emph{critical graph problems}, and \emph{exact optimization
problems}, and they showed some of these problems complete for~$\DP$;
see also the recent
survey~\cite{rie-rot:j:boolean-hierarchies-survey}.
$\DP$ was later generalized by Cai et
al.~\cite{cai-gun-har-hem-sew-wag-wec:j:bh1,cai-gun-har-hem-sew-wag-wec:j:bh2},
who introduced the boolean hierarchy over~$\np$.  
Note that $\DP$ is the second level of this hierarchy.

In his seminal paper, Valiant~\cite{val:j:permanent} initiated the study
of counting problems and introduced the important counting
class~$\sharpp$.  Members of $\sharpp$ are referred to as \emph{$\np$
counting problems}.  A well-known $\np$ counting problem is $\sharpsat$,
the counting version of the satisfiability problem: Given a boolean
formula, how many satisfying assignments does it have?

\begin{definition}[Valiant~\cite{val:j:permanent}]
Let NPTM be a shorthand for nondeterministic polynomial-time Turing
machine.  For any NPTM $M$ and any input~$x$, let $\acc{M}(x)$ denote
the number of accepting computation paths of~$M(x)$, i.e., $\acc{M}$
is a function mapping from $\sigmastar$ to~$\nats$.  Define the
function class $\sharpp = \{ \acc{M} \condition \text{ $M$ is an
NPTM}\}$.
\end{definition}

We now define the notion of \emph{(polynomial-time) parsimonious
reducibility}, which will be used to compare the hardness of solving
$\np$ counting problems.  Intuitively, an $\np$ counting problem $f$
parsimoniously reduces to an $\np$ counting problem $g$ if the
instances of $f$ can be transformed into instances of $g$ such that
the number of solutions of $f$ are preserved under this
transformation.

\begin{definition}
Let $f$ and $g$ be any two given
counting problems mapping from $\sigmastar$ to~$\nats$.  We say
\emph{$f$ (polynomial-time) parsimoniously reduces to $g$} (denoted by
$f \parsimonious g$) if there exists a polynomial-time computable
function $\rho$ such that for each~$x \in \sigmastar$, $f(x) =
g(\rho(x))$.  If $F$ and $G$ are the $\np$ decision problems
corresponding to the $\np$ counting problems $f$ and $g$ with $f
\parsimonious g$, we will also say that $F$ parsimoniously reduces
to~$G$.
\end{definition}

\subsection{Variants of the Tantrix\texttrademark\ Rotation Puzzle Problem
% : Decision, Counting, and Unique Problem
}

\begin{figure}[t!]
  \centering
  \subfigure[Sint]{
    \label{fig:Sint}
\quad
\includegraphics[width=1.5cm]{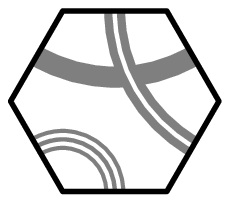}
\quad
  }
  \subfigure[Brid]{
    \label{fig:Brid}
\quad
\includegraphics[width=1.5cm]{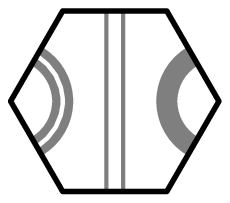}
\quad
  }
  \subfigure[Chin]{
    \label{fig:Chin}
\quad
\includegraphics[width=1.5cm]{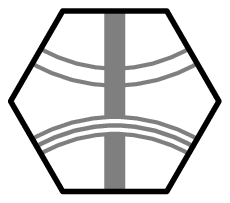}
\quad
  }      
  \subfigure[Rond]{
    \label{fig:Rond}
\quad
\includegraphics[width=1.5cm]{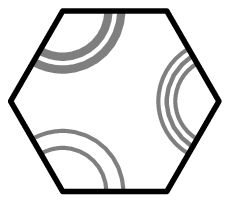}
\quad
  }

  \subfigure[red]{
    \label{fig:red}
\quad
\includegraphics[width=1.5cm]{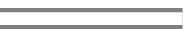}
\quad
  }
  \subfigure[yellow]{
    \label{fig:yellow}
\quad
\includegraphics[width=1.5cm]{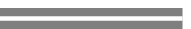}
\quad
  }
  \subfigure[blue]{
    \label{fig:blue}
\quad
\includegraphics[width=1.5cm]{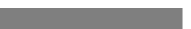}
\quad
  }      
  \subfigure[green]{
    \label{fig:green}
\quad
\includegraphics[width=1.5cm]{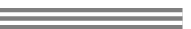}
\quad
  }
  \caption{Tantrix\texttrademark\ tiles and colors}
  \label{fig:TantrixTiles}
\end{figure}

The Tantrix\texttrademark\ rotation puzzle has four kinds of hexagonal
tiles---the \emph{Sint}, the \emph{Brid}, the \emph{Chin}, and the
\emph{Rond}---each of which has three colored lines, where the colors
are chosen among \emph{red}, \emph{yellow}, \emph{blue}, and
\emph{green}, see Figure~\ref{fig:TantrixTiles}(a)--(d).  This gives a
total of 56 different tiles.  Since we aren't using actually colored
figures, we encode the colors as shown in
Figure~\ref{fig:TantrixTiles}(e)--(h).

Holzer and Holzer~\cite{hol-hol:j:tantrix} showed that the decision
problem Tantrix\texttrademark\ rotation puzzle (which we denote
by~$\trp$, for short) is $\np$-complete.  In this paper, we introduce
and study~$\sharptrp$, the counting version of~$\trp$.

\begin{figure}[h!]
  \centering
\input{coordinate-system-4color.eepic}
  \caption{A two-dimensional hexagonal coordinate system}
  \label{fig:coordinate-system}
\end{figure}
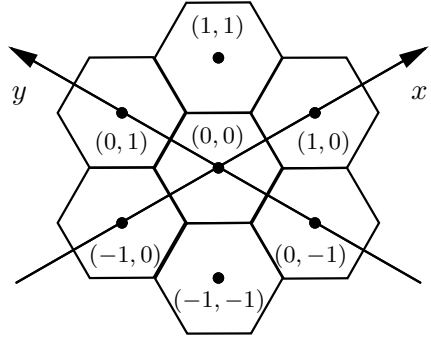

We now briefly describe the formalism introduced by Holzer and Holzer
\cite{hol-hol:j:tantrix} to define~$\trp$, since the same formalism is
useful for defining~$\sharptrp$.  In particular, to represent the
instances of both these problems, a two-dimensional hexagonal
coordinate system is used, see Figure~\ref{fig:coordinate-system}.  In
this system, two distinct pairs $a=(u,w)$ and $b=(v,x)$ from
$\mathbb{Z}^2$
% , where $a \neq b$,
are adjacent if and only if
% either
one of the following four conditions is
satisfied:
\begin{enumerate}
\item $u=v$ and $|w-x|=1$,
\item $|u-v|=1$ and $w=x$,
\item $u-v=1$ and $w-x=1$, and
\item $u-v=-1$ and $w-x=-1$.
\end{enumerate}

Let $T$ be the set of all Tantrix\texttrademark\ tiles.  Let
$\mathcal{A}$ be a (partial)
function mapping the elements of
$\mathbb{Z}^2$ to~$T$,
i.e., for those $v \in \mathbb{Z}^2$ on which $\mathcal{A}$ is defined,
$\mathcal{A}(v)$ is the type of the tile located at~$v$.
The set $\shape(\mathcal{A}) = \{ v \in \mathbb{Z}^2 \condition
\mathcal{A}(v) \text{ is defined}\}$ gives the positions in
$\mathbb{Z}^2$ at which tiles are placed.
% in the two-dimensional hexagonal space.
For all $a,b \in \shape(\mathcal{A})$, $\mathcal{A}(a)$ is adjacent to
$\mathcal{A}(b)$ if and only if $a$ is adjacent to~$b$.

$\trp$ is then defined as follows (note that the
initial orientation is not specified, as it doesn't matter for the
question of whether the decision problem $\trp$ is
solvable)~\cite{hol-hol:j:tantrix}:\footnote{As noted
by Holzer and Holzer~\cite{hol-hol:j:tantrix}, there is a difference
between their definition of $\trp$, which allows holes  in $\trp$ instances,
and the original Tantrix\texttrademark\
game, which does not allow holes. 
% means that the graph whose vertices correspond to tiles and whose
% edges correspond to adjacencies between tiles need not be connected.
The problem of whether the analog of
$\trp$ \emph{without} holes still is $\np$-complete is open.
}
{ \samepage
\begin{desctight}
\item[Name:] Tantrix\texttrademark\ Rotation Puzzle ($\trp$,
for short).

\item[Given:] A finite shape function $\mathcal{A} : \mathbb{Z}^2
\rightarrow T$, appropriately encoded as a string.
% in~$\sigmastar$.

\item[Question:] Is the rotation puzzle defined by $\mathcal{A}$
solvable, i.e., does there exist a rotation
of the given tiles at their positions such that at each joint edge of two
adjacent tiles the corresponding colors match?
\end{desctight}
} % ends \samepage

\begin{figure}[htbp]
  \centering
  \subfigure[Puzzle]{
    \label{fig:bsp-puzzle}
\includegraphics[width=2.8cm]{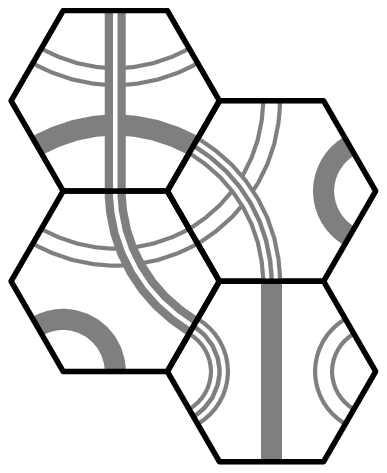}
\quad
  }
  \subfigure[Solution]{
    \label{fig:bsp-solution}
\quad
\includegraphics[width=2.8cm]{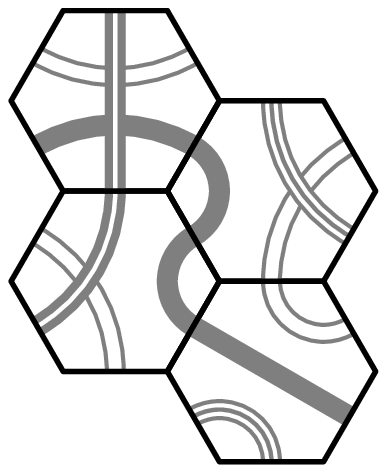}
  }
  \caption{An example of a $\trp$ instance and its solution}
  \label{fig:Bsp}
\end{figure}

For any given $\trp$ instance~$\mathcal{A}$, a \emph{solution
of~$\mathcal{A}$} is a specification (in some appropriate encoding) of
each tile in $\shape(\mathcal{A})$ in some particular orientation such
that for each joint edge of two adjacent tiles the corresponding
colors match.  Figure~\ref{fig:Bsp} gives an example of a rotation
puzzle instance and its solution.  Let $\solutiontrp{\mathcal{A}}$
denote the set of solutions of a given $\trp$ instance~$\mathcal{A}$.
So $\mathcal{A}$ is in $\trp$ (viewed as a language) if and only if
the set $\solutiontrp{\mathcal{A}}$ is nonempty.

We now define the counting version and the unique version of~$\trp$,
which will be considered in
Sections~\ref{sec:sharptrp-is-sharpp-complete}
and~\ref{sec:uniquetrp-is-dp-complete}.

\begin{definition}
\label{def:sharptrp}
\begin{enumerate}
\item The \emph{Tantrix\texttrademark\ rotation puzzle counting
problem} is the function $\sharptrp : \sigmastar \rightarrow
\nats$ defined by
\[
\sharptrp(\mathcal{A}) = \| \solutiontrp{\mathcal{A}} \|,
\]
where we assume that inputs $\mathcal{A}$ are appropriately encoded as
strings in $\sigmastar$ and
% outputs
function values are nonnegative integers (represented in binary).

\item The \emph{unique Tantrix\texttrademark\ rotation puzzle problem}
is defined by
\[
\utrp = \{\mathcal{A} \condition \sharptrp(\mathcal{A}) = 1 \}.
\]
\end{enumerate}
\end{definition}

\section{Satisfiability Parsimoniously Reduces to the 
Tantrix\texttrademark\ Rotation Puzzle Problem}
\label{sec:sharptrp-is-sharpp-complete}

Our main result is Theorem~\ref{thm:sharptrp-is-sharpp-complete} the
proof of which will be presented in Sections~\ref{sec:wire}
% , \ref{sec:gate}, \ref{sec:input-output}, and
through~\ref{sec:proof}.

\begin{theorem}
\label{thm:sharptrp-is-sharpp-complete}
$\sharpsat \parsimonious \sharptrp$.
\end{theorem}

To prove $\trp$ $\np$-complete, Holzer and
Holzer~\cite{hol-hol:j:tantrix} gave a reduction from the
$\np$-complete problem $\circuitandnotsat$ (see
Cook~\cite{coo:c:theorem-proving}), which is defined as follows.

{\samepage
\begin{desctight}
\item[Name:] $\circuitandnotsat$.

\item[Given:] A boolean circuit $C$ with AND and NOT gates.

\item[Question:] Does there exist a truth assignment to the input
gates of $C$ such that $C$ under this assignment evaluates to
\emph{true}?
\end{desctight}
} % ends \samepage

% Figure~\ref{fig:circuit} gives an example of such a circuit.
Holzer and Holzer's construction simulates the computation of such a
boolean circuit $C$ by a Tantrix\texttrademark\ rotation puzzle such
that $C$ evaluates to \emph{true} for some assignment to its variables if and
only if the puzzle has a solution.

Our definition of boolean circuits follows Holzer and
Holzer~\cite{hol-hol:j:tantrix}, who view a boolean circuit $C$ with
input variables $x_1, x_2, \ldots , x_n$ as a sequence $(\alpha_1,
\alpha_2, \ldots ,\alpha_m)$ of steps such that the $i$th instruction
$\alpha_i$ has one of the following forms:
\begin{enumerate}
\item for each $i$ with $1 \leq i \leq n$, $\alpha_i = x_i$,
\item for each $i$ with $n+1 \leq i \leq m$, either $\alpha_i =
\mbox{AND}(j,k)$ or $\alpha_i = \mbox{NOT}(j)$, where $j\leq k < i$.
\end{enumerate}
Depending on the truth values of the input variables the output gate
evaluates to \emph{true} or \emph{false} in the standard way.

\begin{figure}[h!]
  \centering
\includegraphics[width=4cm]{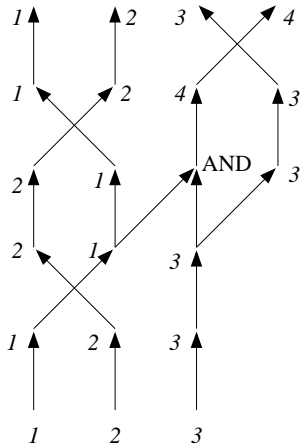}
  \caption{Example of a circuit following Goldschlager's
  transformation~\cite{gol:j:monotone-planar-circuits}}
  \label{fig:circuit-bsp}
\end{figure}

% DB: Das habe ich neu fuer das cross eingefuegt, vielleicht faellt dir
% ja noch was besseres ein
In general, circuits can contain
% arbitrary 
any number of wire crossings, which cannot
easily be realized by Tantrix\texttrademark\ subpuzzles.
To build a circuit containing only crossings of two neighboring wires,
Holzer and Holzer follow Goldschlager's 
procedure~\cite{gol:j:monotone-planar-circuits}: If
$\alpha_i = \mbox{AND}(j,k)$, move wire $j$ immediately
to the left of~$k$, put an AND gate in for $\alpha_i$, move wire $j$
back to its starting point, and finally move wire $i$ to the far right. The
cases of instruction $\alpha_i$ being either NOT$(j)$ or $x_i$ are treated in
a similar way.  Figure~\ref{fig:circuit-bsp} shows a part of a circuit
with wire crossings, which computes $\alpha_4=\mbox{AND}(1,3)$.
Obviously, there are only crossings of two directly
% neighboured
adjacent wires. 
Note that this transformation from general to almost planar circuits
can be done in deterministic logarithmic space.

Now, to build a Tantrix\texttrademark\ Rotation Puzzle that simulates
such a circuit, Holzer and Holzer 
use a truly planar ``cross-over''
% circuit 
gadget that was proposed by McColl~\cite{mcc:j:planar-crossovers}.
McColl's circuit gadget uses boolean AND and NOT gates to simulate
the crossings of any two 
adjacent wires.  Each such ``cross-over'' gadget needs a total of
$14$ instruction steps 
and involves
% $12$
twelve AND and
% $9$
nine NOT gates.  Since many crossings can occur
in the originally given circuit, this may lead to a considerable (albeit
still polynomial) blow-up of the Tantrix\texttrademark\ puzzle constructed
in our reduction.  That is why we propose, as a more efficient
alternative to the reduction presented in our previous
paper~\cite{bau-rot:c:tantrix},
to simulate these cross-overs directly via a new Tantrix\texttrademark\
subpuzzle, the CROSS subpuzzle presented in Figure~\ref{fig:cross-par-4trp}.
Unlike the reductions given
in~\cite{hol-hol:j:tantrix,bau-rot:c:tantrix}, via such CROSS
subpuzzles our reduction doesn't need the transformation from general to
planar circuits that requires many additional gates and
instruction steps caused by wire crossings.

% DB: Ich weiss nicht mehr welches Beispiel in HH04 war (und habe es auch
% nicht hier zum nachgucken)
% Oder besser das groessere Beispiel vor und nach der Transformation?
% \begin{figure}[t!]
%   \centering
% \includegraphics[width=4.5cm]{../Bilder-EPS-sw/circuit-bsp.eps}
%   \caption{Example for a circuit following Goldschlager's
%   transformation~\cite{gol:j:monotone-planar-circuits}}
%   \label{fig:circuit-bsp}
% \end{figure}

Furthermore, our construction will modify Holzer and Holzer's
reduction~\cite{hol-hol:j:tantrix} in such a way that there is a
one-to-one correspondence between the solutions of the given
$\circuitandnotsat$ instance and the solutions of the resulting
rotation puzzle instance; hence our reduction is parsimonious. 
% The reduction
%employs planar cross-over gates (consisting of AND and NOT gates only)
%to avoid wire crossings of the given circuit; for technical details
%and examples, see~\cite{hol-hol:j:tantrix}.

To simulate the circuit by a rotation puzzle, a number of subpuzzles
are used.  The color blue in these subpuzzles will represent the truth
value \emph{true}, and the color red will represent \emph{false}.
This color encoding at the inputs and outputs of the subpuzzles thus
represent the truth values of the circuit's gates and wires.\\

In the following sections, we present our modified subpuzzles and, to
allow comparison, we also present Holzer and Holzer's
subpuzzles~\cite{hol-hol:j:tantrix}.
To indicate the differences between their original and our modified
subpuzzles, tiles with more than one possible
solution in the original subpuzzles
will have a grey instead of a black border, and 
we highlight all modified tiles in our new subpuzzles by a 
grey  instead of a white background (unless stated otherwise).

Another difference between our proof here and the proofs of Holzer and
Holzer~\cite{hol-hol:j:tantrix} and a preliminary version of this
paper~\cite{bau-rot:c:tantrix} regards the analysis of the subpuzzles.
In particular, we will here focus on the color sequences of the
various Tantrix\texttrademark\ tiles.  This will allow us to give the
arguments more
% succinctly and more elegantly
tersely.  For example, the
tile in Figure~\ref{fig:Rond} has the clockwise color sequence
%%% JR: Ich glaube, Fig. 1(d) hat eine andere color sequence, siehe unten.
% ${\tt yellow}-{\tt blue}-{\tt yellow}-{\tt green}-{\tt green}-{\tt blue}$,
${\tt yellow}$-${\tt green}$-${\tt green}$-${\tt red}$-${\tt red}$-${\tt yellow}$,
which will be abbreviated as
% ${\tt ybyggb}$
${\tt yggrry}$.
% In what
% follows, the modified subpuzzles are presented.  To allow comparison,
% the original subpuzzles from~\cite{hol-hol:j:tantrix} are presented in
% the appendix.

\subsection{Wire Subpuzzles}
\label{sec:wire}

Wires of the circuit are simulated by the subpuzzles WIRE, MOVE,
COPY, and CROSS.

\begin{figure}[h!]
  \centering
  \subfigure[In: \emph{true}]{
    \label{fig:wire-4trp-t}
    \quad
    \quad
    \includegraphics[height=2.4cm]{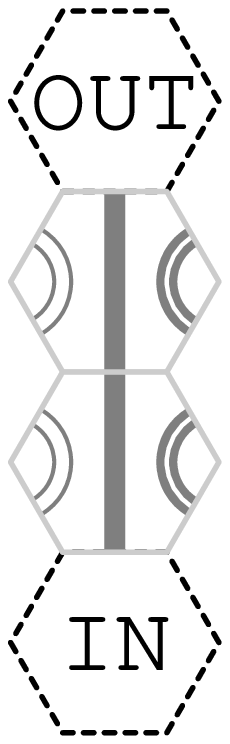}%\hspace{1cm}
    \quad
    \quad
  }
  \quad
  \subfigure[In: \emph{false}]{
    \label{fig:wire-4trp-f}
    \quad
    \quad
    \includegraphics[height=2.4cm]{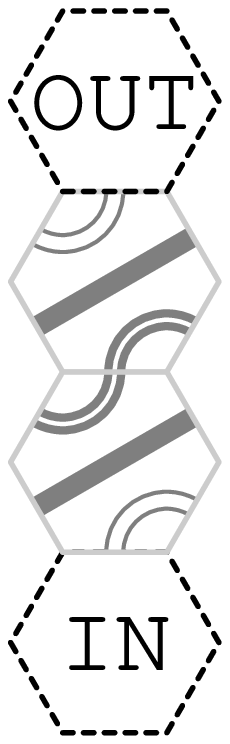}%\hspace{1cm}
    \quad
    \quad
  }
  \quad
  \subfigure[Scheme]{
    \label{fig:wire-4trp-s}
    \quad
    \quad 
    \includegraphics[height=2.4cm]{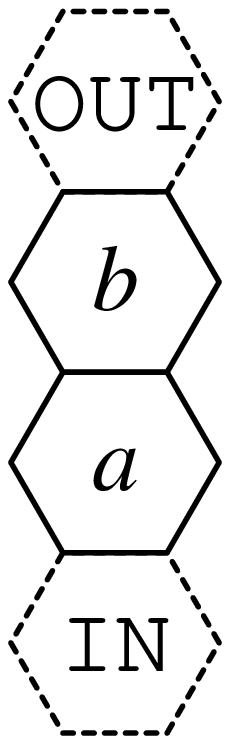}
    \quad
    \quad 
  } 
  \caption{Original subpuzzle WIRE, see~\cite{hol-hol:j:tantrix}}
  \label{fig:wire-4trp}
\end{figure}

\begin{figure}[h!]
  \centering
  \subfigure[In: \emph{true}]{
    \label{fig:wire-par-4trp-t}
    \quad 
    \quad 
    \includegraphics[height=2.4cm]{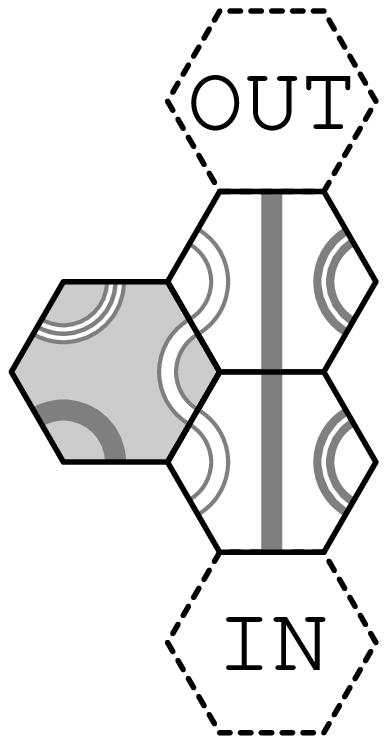}
    \quad 
    \quad 
  }
  \subfigure[In: \emph{false}]{
    \label{fig:wire-par-4trp-f}
    \quad 
    \quad 
    \includegraphics[height=2.4cm]{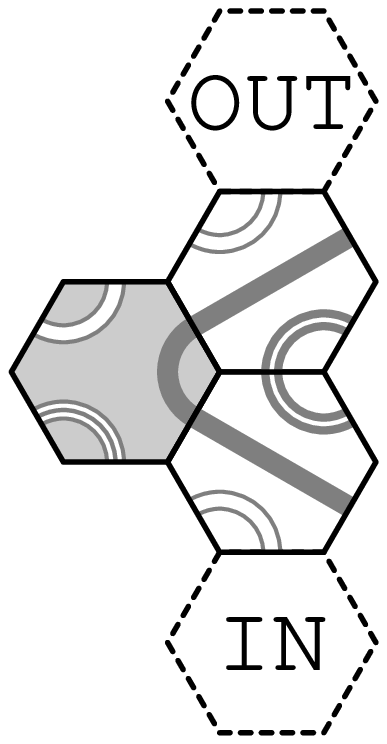}
    \quad 
    \quad 
  }
  \subfigure[Scheme]{
    \label{fig:wire-par-4trp-s}
    \quad
    \quad 
    \includegraphics[height=2.4cm]{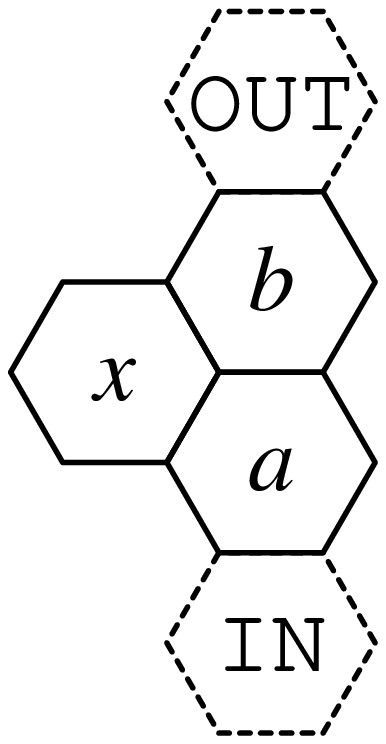}
    \quad
    \quad 
  } 
  \caption{Modified subpuzzle WIRE}
  \label{fig:wire-par-4trp}
\end{figure}

To simulate simple vertical wires, the WIRE subpuzzle is
% employed
used.
The original version of Holzer and Holzer~\cite{hol-hol:j:tantrix} is
presented in Figure~\ref{fig:wire-4trp}. It is easy to see that both tiles
have two possible orientations for each input color.
We modify this subpuzzle as shown in Figure~\ref{fig:wire-par-4trp},
inserting a new Rond at position~$x$.  Without this tile, the possible color
sequences for the edges of tile $x$ joint with $a$ and $b$ are ${\tt rr}$, 
${\tt ry}$, ${\tt yr}$, and ${\tt yy}$ if the input color is \emph{blue}, 
and are ${\tt bb}$, ${\tt by}$, ${\tt yb}$, and ${\tt yy}$ if the input color 
is \emph{red}.  However, since the new tile $x$ does not contain the color
\emph{yellow}, the solutions are fixed with, respectively,
$bb$ and $rr$ at the joint edges
of tile $x$ with $a$ and $b$, and we obtain unique solutions for both 
input colors.

To build longer wires, several WIRE subpuzzles are connected.  Note that 
this WIRE subpuzzle has height two. This forces all other subpuzzles to have
even height, because they must be connected by WIRE subpuzzles.\\

% 
% Figure~\ref{fig:wire-par-4trp} shows the modified WIRE subpuzzle, which simply
% represents a vertical wire.  Longer wires can be built by using
% several WIRE subpuzzles.  A single WIRE has height two, which implies
% that all other subpuzzles must have even height.  (Otherwise it
% wouldn't be possible to simulate a circuit by a rotation puzzle.)  It
% is easy to see that the original WIRE subpuzzle
% from~\cite{hol-hol:j:tantrix} (see Figure~\ref{fig:wire-4trp}) 
% %in the appendix) 
% has more than one valid solution with the input colors
% blue and red.  In particular, tile $a$ and tile $b$ have two possible
% orientations for each input color, so there are four possible
% solutions.
% However, by inserting a \emph{Rond} in the colors blue, red, and green
% at position~$x$, we obtain a unique solution.  If the input color is
% blue, there is a blue vertical line.  Tiles $a$ and $b$ now must have
% red at the edge adjacent to tile~$x$, since $x$ doesn't have yellow.
% If the input color is red, tiles $a$ and $b$ have a choice between
% either blue or yellow for the edge joint with~$x$.  Again, since $x$
% doesn't have yellow, the solution is unique.

\begin{figure}[h!]
  \centering
  \subfigure[In: \emph{true}]{
    \label{fig:move-4trp-t}
    \quad
    \includegraphics[height=3.6cm]{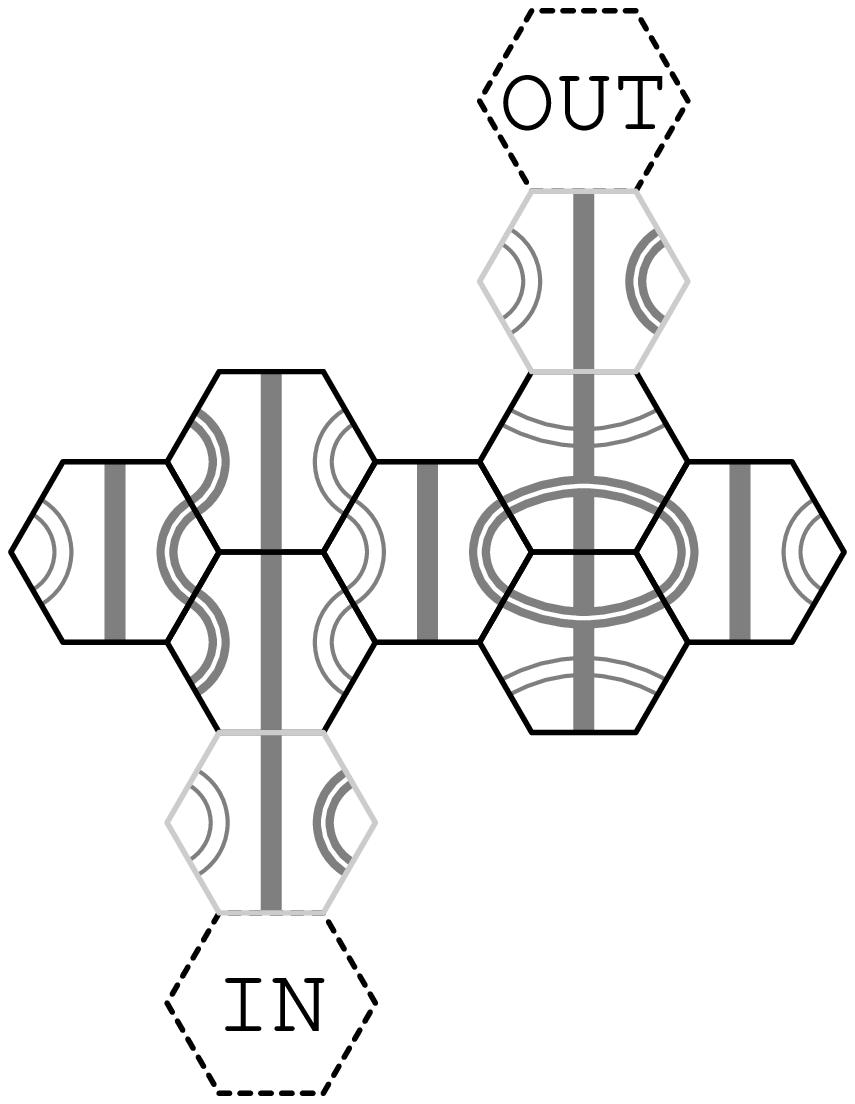}
    \quad
  }
  \subfigure[In: \emph{false}]{
    \label{fig:move-4trp-f}
    \quad
    \includegraphics[height=3.6cm]{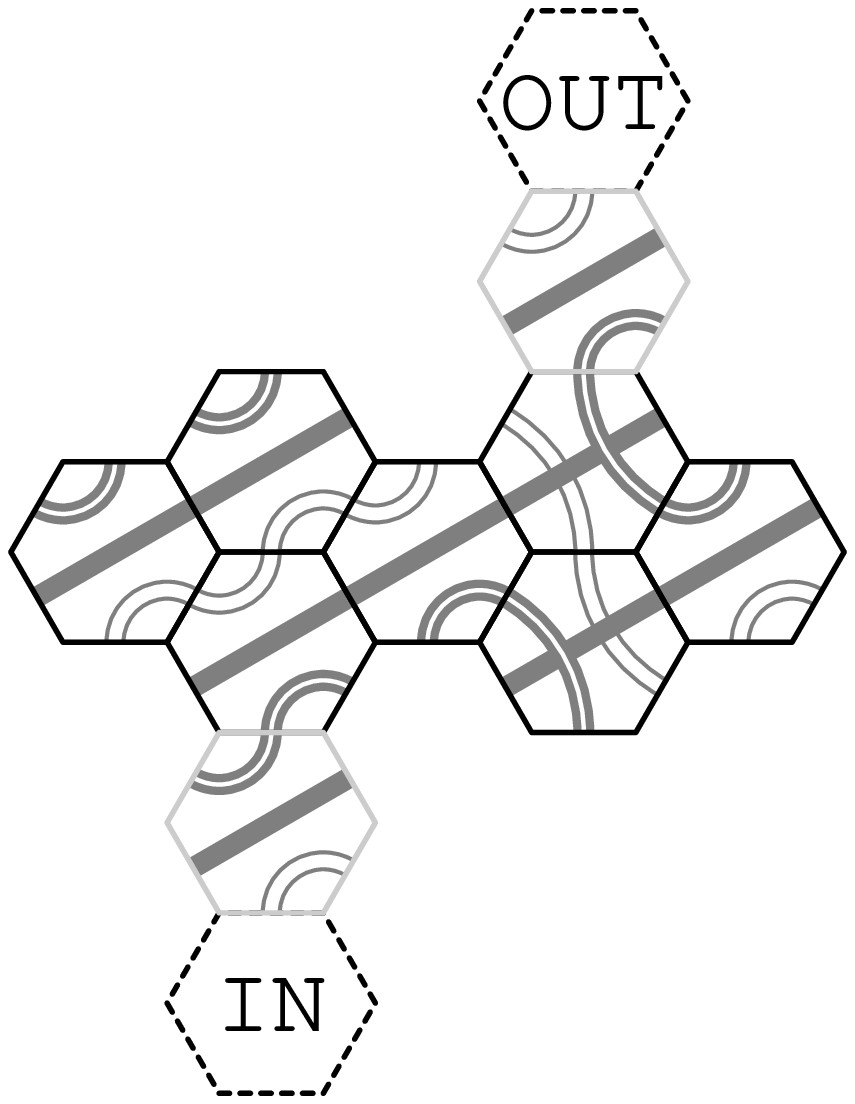}
    \quad
  }  
  \subfigure[Scheme]{
    \label{fig:move-4trp-s}
    \quad
    \includegraphics[height=3.6cm]{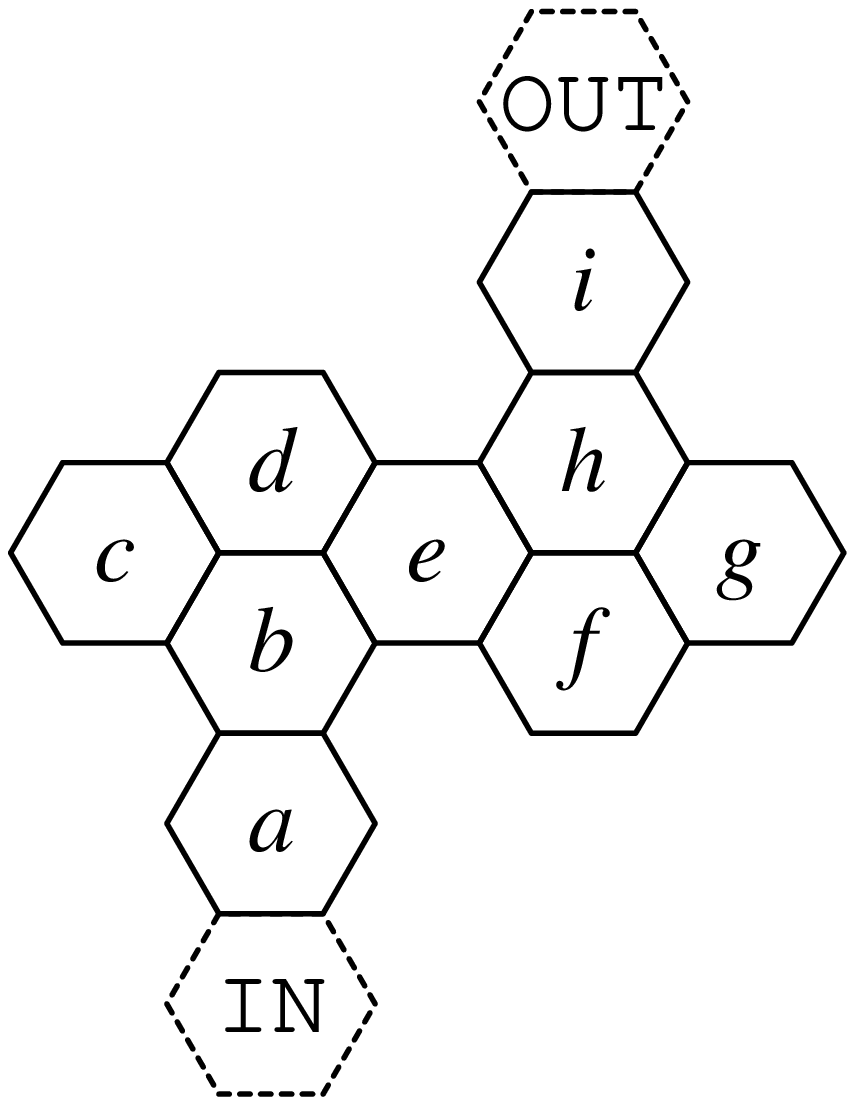}
    \quad 
  }    
  \caption{Original subpuzzle MOVE, see~\cite{hol-hol:j:tantrix}}
  \label{fig:move-4trp}
\end{figure}

\begin{figure}[h!]
  \centering
  \subfigure[In: \emph{true}]{
    \label{fig:move-par-4trp-t}
    \quad 
    \includegraphics[height=3.6cm]{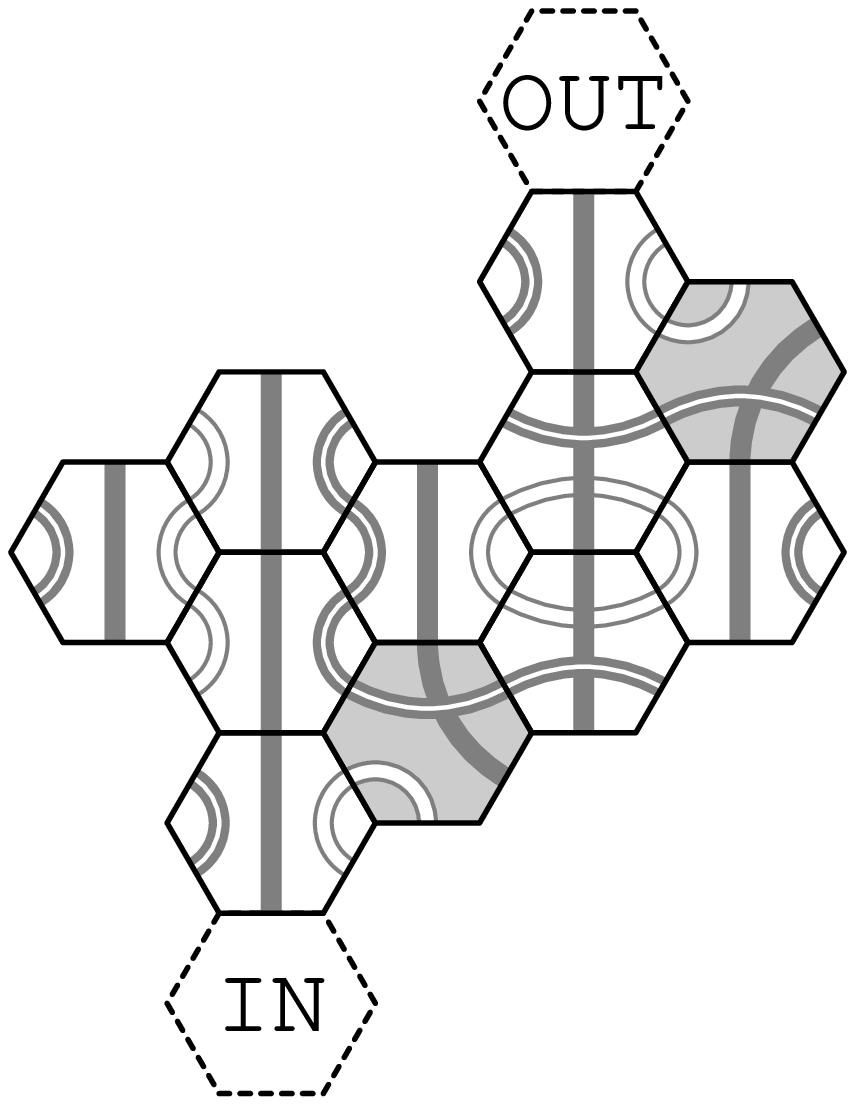}
    \quad 
  }
  \subfigure[In: \emph{false}]{
    \label{fig:move-par-4trp-f}
    \quad 
    \includegraphics[height=3.6cm]{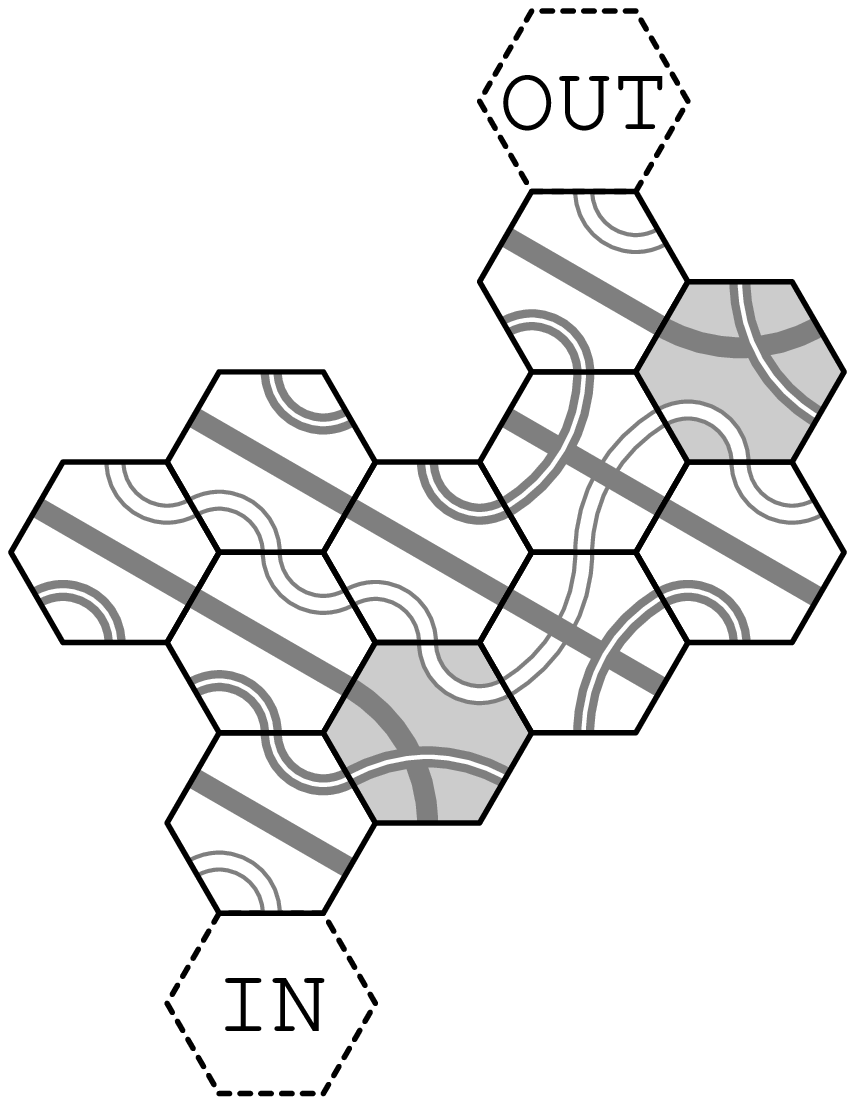}
    \quad 
  }
  \subfigure[Scheme]{
    \label{fig:move-par-4trp-s}
    \quad
    \includegraphics[height=3.6cm]{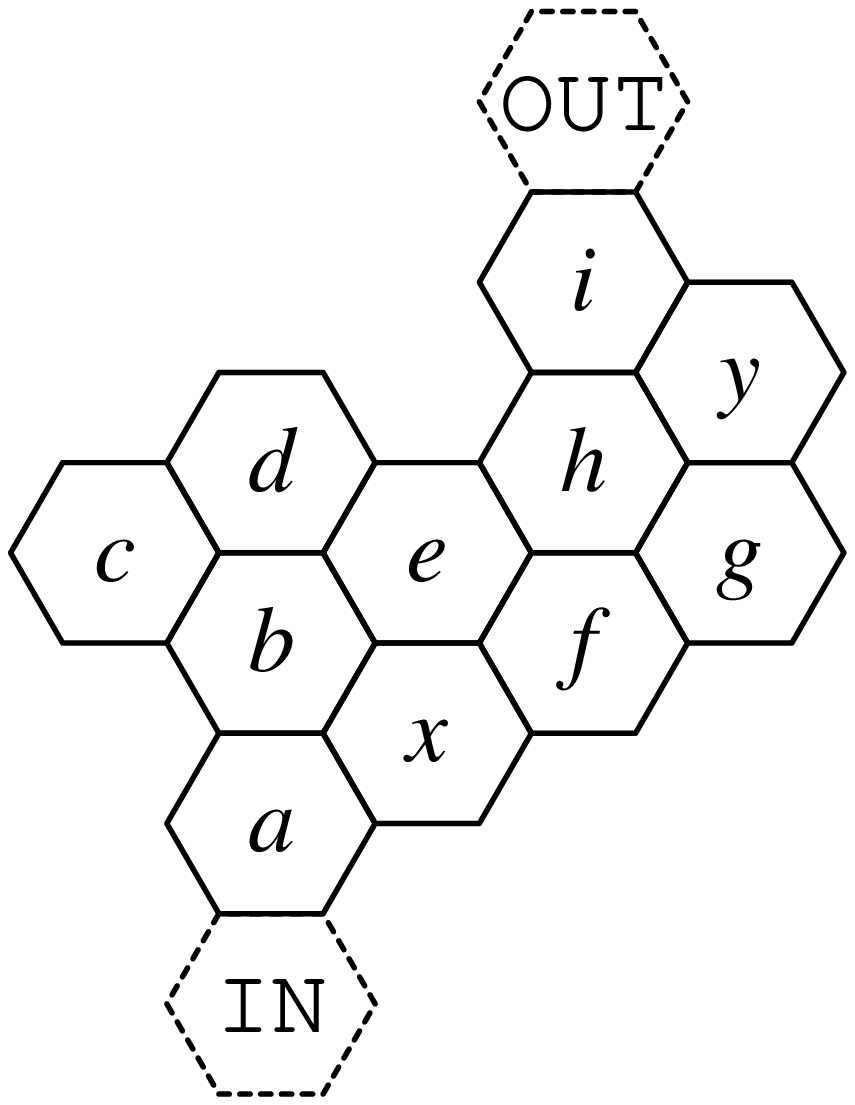}
    \quad 
  }    
  \caption{Modified subpuzzle MOVE}
  \label{fig:move-par-4trp}
\end{figure}

By the MOVE subpuzzle the circuit's wires can be moved one position to
the left or one position to the right.  We discuss a move to the right
in detail and mention that a move to the left can be handled
analogously.  The original version of Holzer and
Holzer~\cite{hol-hol:j:tantrix} is shown in
Figure~\ref{fig:move-4trp}, and our modified subpuzzle is shown in
Figure~\ref{fig:move-par-4trp}.

In the original MOVE subpuzzle, tiles $a$ and $i$ are marked because
they have two orientations for the solutions shown in
Figure~\ref{fig:move-4trp}.  Note, however, that in addition to
the ambiguity caused by tiles $a$ and $i$, there does exist still
another solution for each input color.  In particular, if the input
color is \emph{blue} then one can simply swap the colors \emph{red}
and \emph{yellow} to obtain another solution, and if the input color
is \emph{red} then tiles $b$ and $d$ can be rotated by $60$ and all
other tiles by $240$ degrees in clockwise direction.

We fix the solution by inserting tiles $x$ and~$y$. First, consider
the case that the input color is \emph{blue}. It is clear that the
edge of tile $e$ joint with tile $x$ must be \emph{blue}, and since
the edges of tiles $x$ joint with tiles $b$ and $f$ must have the same
color, the orientation of tile $x$ is fixed. This fixes also the
orientation of all other tiles except $a$ and~$i$. The orientation of
tile $a$ is fixed with \emph{red} at the edge joint with~$x$. Tile $y$
fixes the orientation of tile~$i$, since it does not contain the color
sequence ${\tt byy}$, but the color sequence ${\tt byr}$ for the edges
joint with tiles~$g$, $h$, and~$i$.  The case of \emph{red} being the
input color can be handled similarly.

\begin{figure}[h!]
  \centering
  \subfigure[In: \emph{true}]{
    \label{fig:copy-4trp-t}
    \quad
    \includegraphics[height=3.6cm]{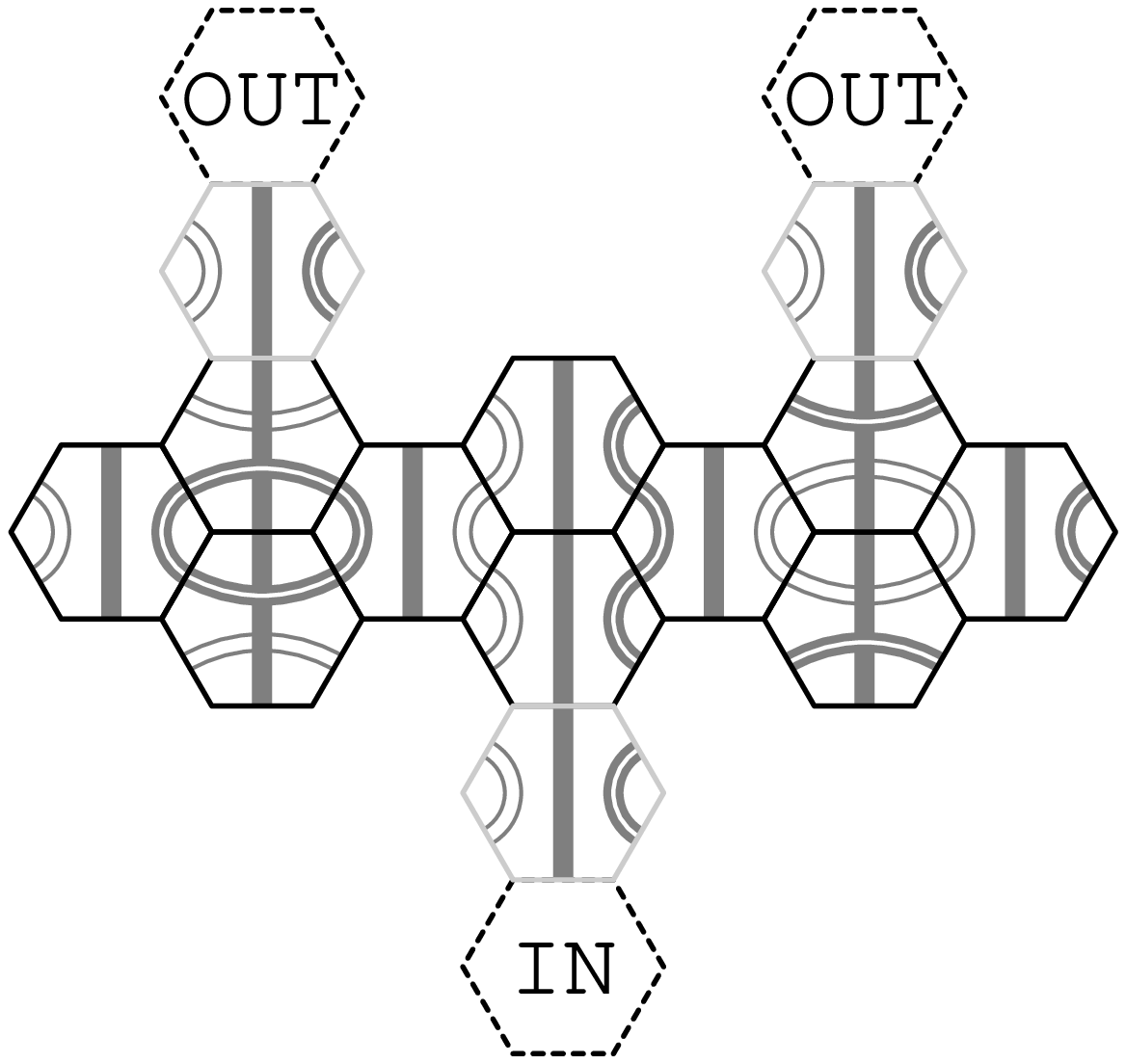}
    \quad
  }
  \subfigure[In: \emph{false}]{
    \label{fig:copy-4trp-f}
    \quad
    \includegraphics[height=3.6cm]{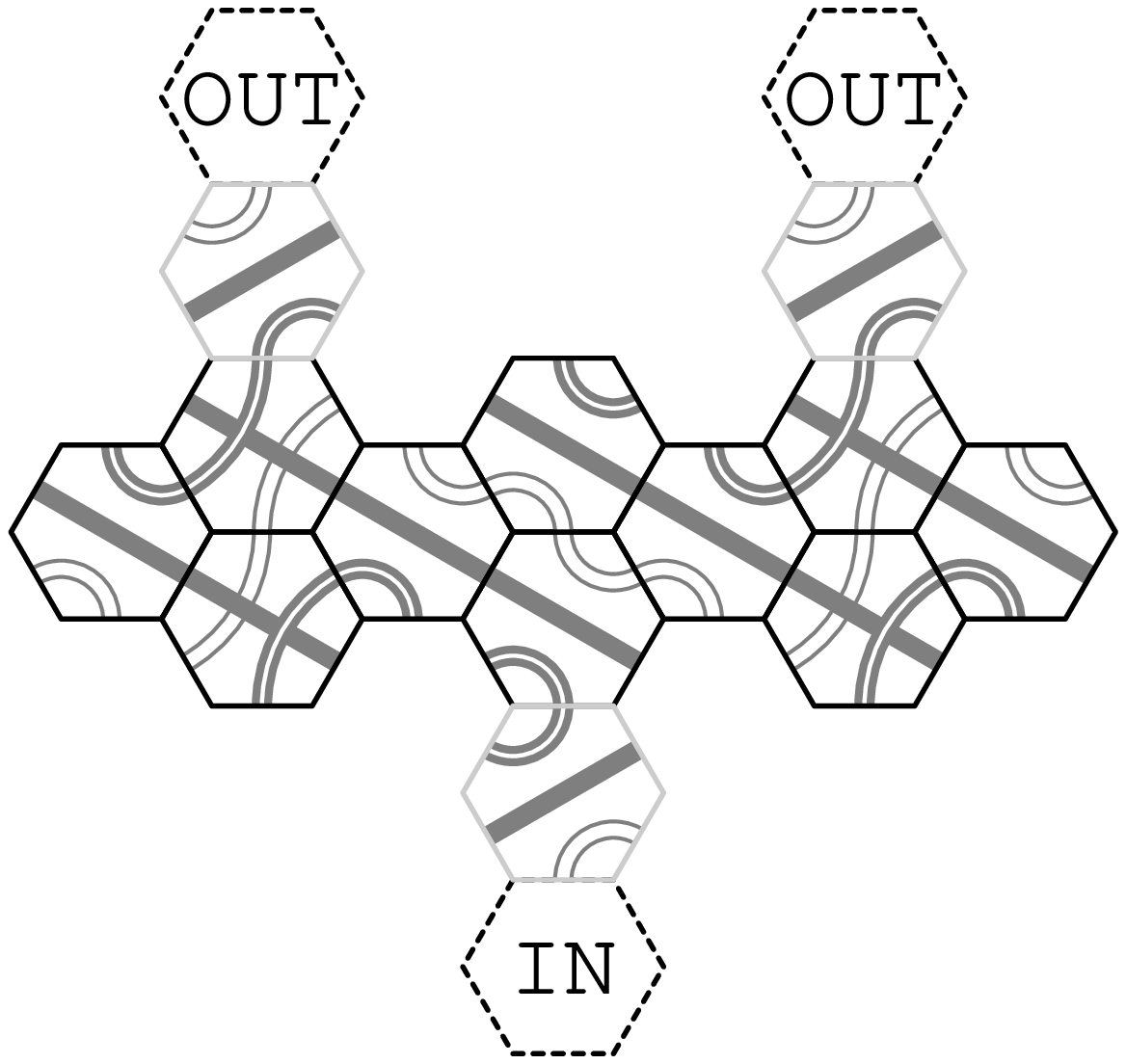}
    \quad
  }   
  \subfigure[Scheme]{
    \label{fig:copy-4trp-s}
    \quad 
    \includegraphics[height=3.6cm]{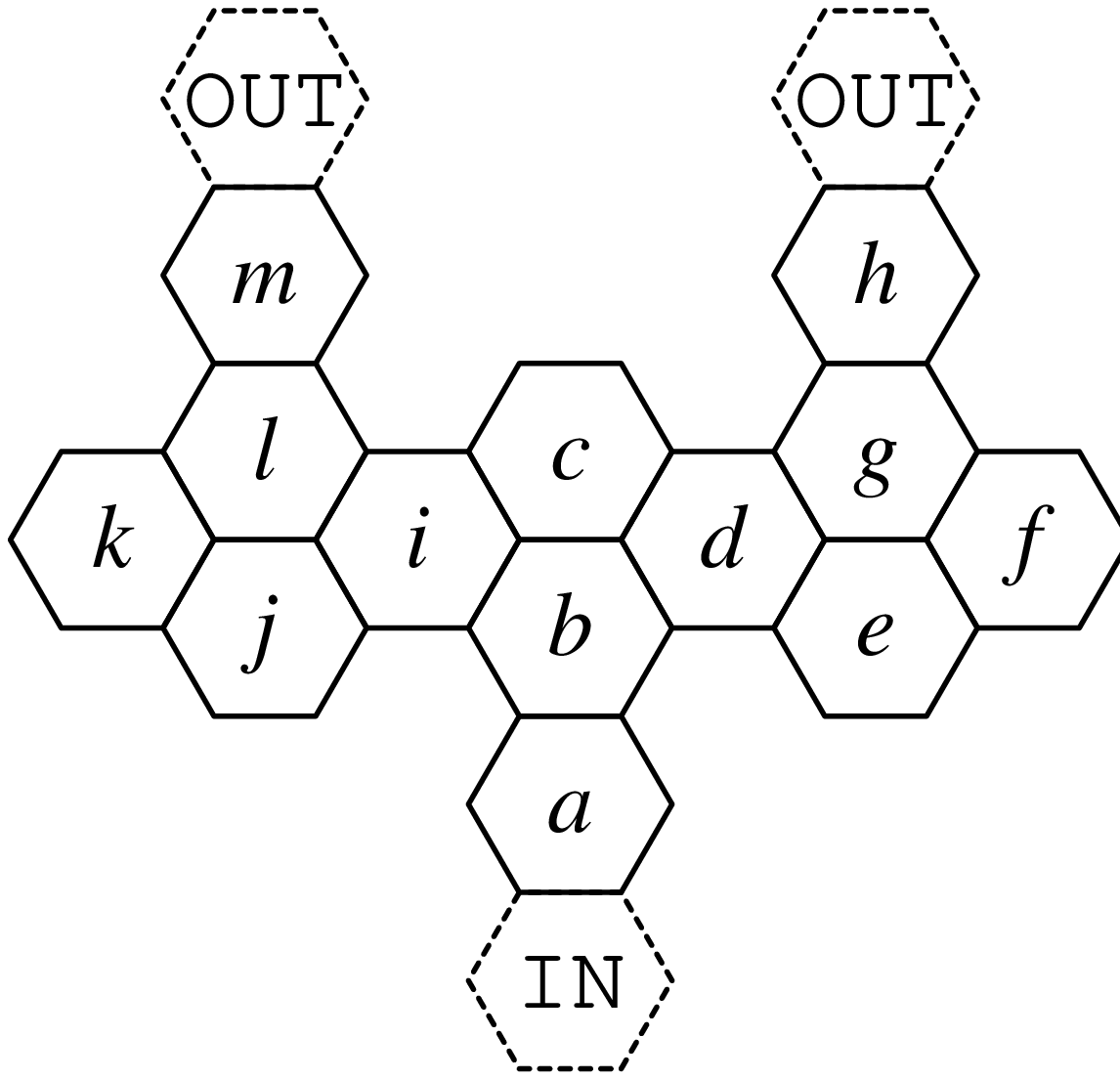}
    \quad
  }
  \caption{Original subpuzzle COPY, see~\cite{hol-hol:j:tantrix}}
  \label{fig:copy-4trp}
\end{figure}

\begin{figure}[h!]
  \centering
  \subfigure[In: \emph{true}]{
    \label{fig:copy-par-4trp-t}
    \quad
    \includegraphics[height=3.6cm]{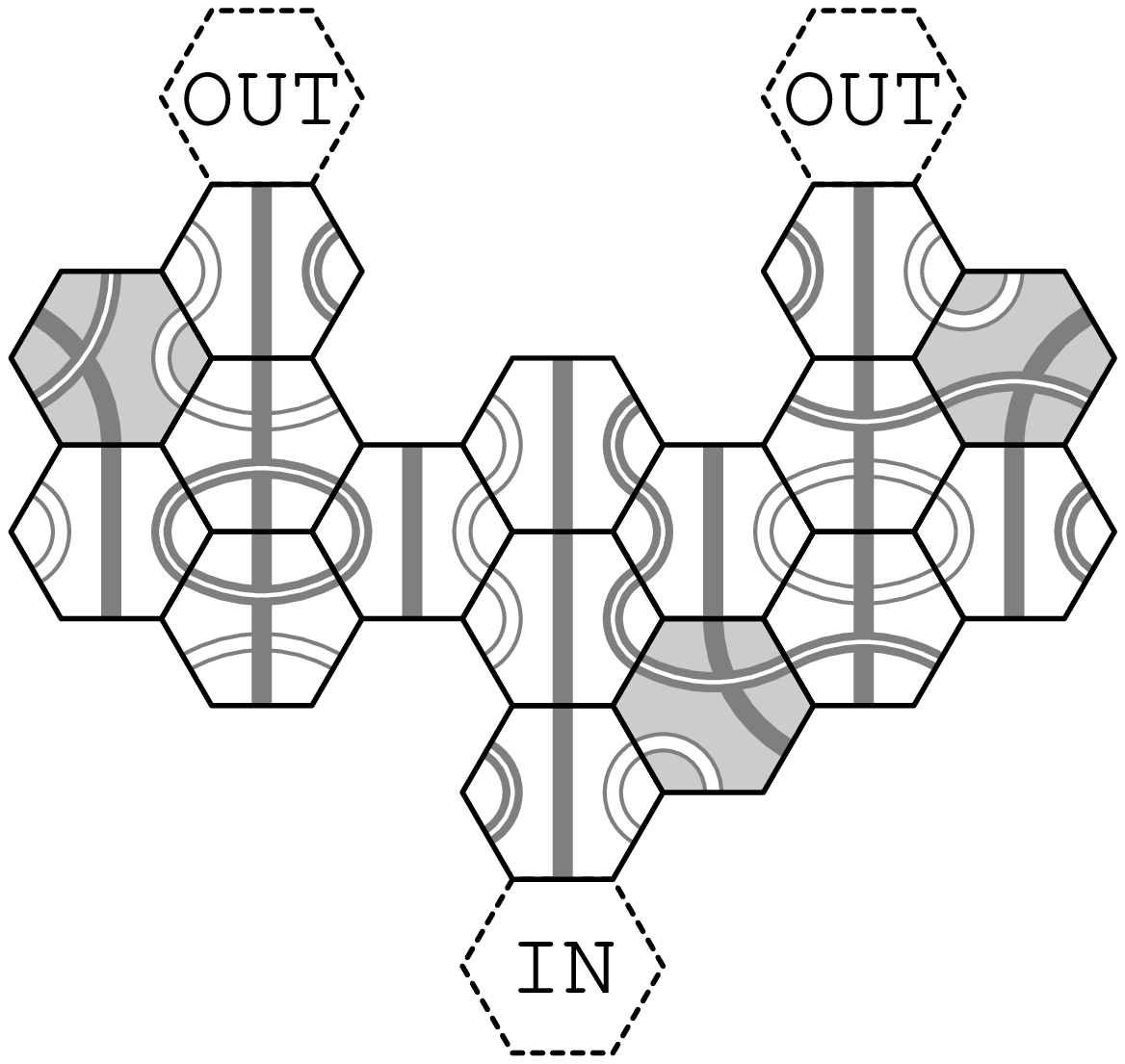}
    \quad 
  }
  \subfigure[In: \emph{false}]{
    \label{fig:copy-par-4trp-f}
    \quad 
    \includegraphics[height=3.6cm]{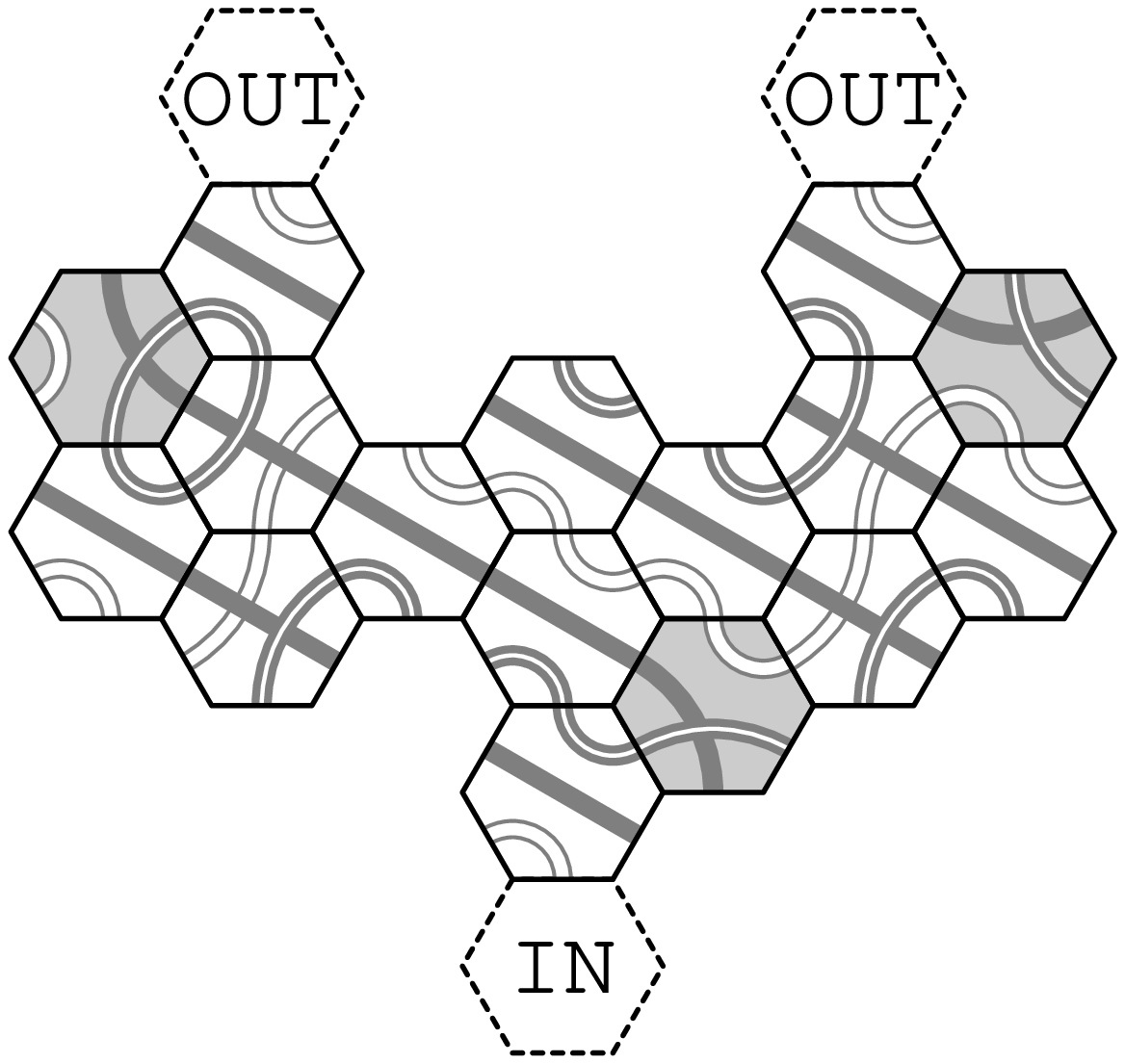}
    \quad 
  }
  \subfigure[Scheme]{
    \label{fig:copy-par-4trp-s}
    \quad 
    \includegraphics[height=3.6cm]{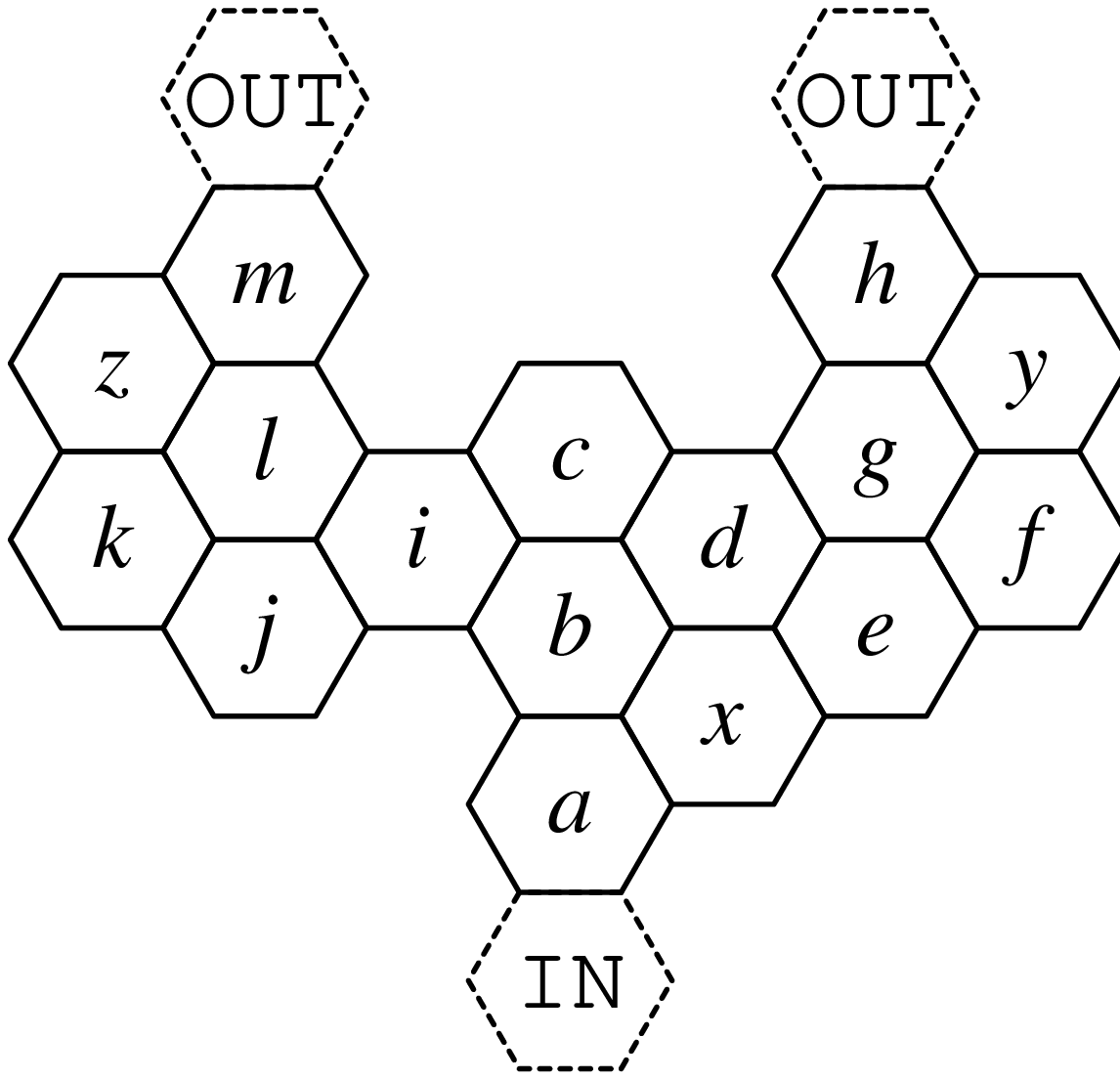}
    \quad
  }
  \caption{Modified subpuzzle COPY}
  \label{fig:copy-par-4trp}
\end{figure}

The COPY subpuzzle is used to ``split'' a wire into two copies.  Its
original version from~\cite{hol-hol:j:tantrix} is shown in
Figure~\ref{fig:copy-4trp}, and the modified version is shown in
Figure~\ref{fig:copy-par-4trp}.  In its original version, tiles $a$
through $i$ are a move to the right, merged with a move to the left
consisting of tiles $a$ through $d$ and tiles $i$ through~$m$.  To
obtain a unique solution, we insert the same tiles $x$ and $y$ as for
the MOVE subpuzzle. By the same argument as above, this fixes the
orientation of all other tiles, except tile $m$. But inserting tile
$z$, which is of the same type as tile~$y$, also fixes the orientation
of~$m$.  Thus, the solution for this subpuzzle is also unique for both
input colors. \\

% 
% Finally, Figure~\ref{fig:copy-par-4trp} presents the modified COPY subpuzzle,
% which can be used to ``split'' a wire into two copies.  Its structure
% is akin to the MOVE subpuzzle, though it is wider and has two outputs.
% Due to symmetry, the original COPY subpuzzle
% from~\cite{hol-hol:j:tantrix} (see Figure~\ref{fig:copy-4trp})
% % in
% %the appendix) 
% has again more than one valid solution.  However, we can
% enforce a unique solution (for both input colors, blue and red) by
% inserting three asymmetric \emph{Sint} tiles in the colors red, blue,
% and yellow at positions~$x$, $y$, and~$z$.  The argument then is similar
% to the one for the MOVE subpuzzle.

\begin{figure}[h]
  \centering
  \subfigure[In: \emph{true, true}]{
    \label{fig:cross-par-4trp-tt}
    \quad
    \includegraphics[height=4.8cm]{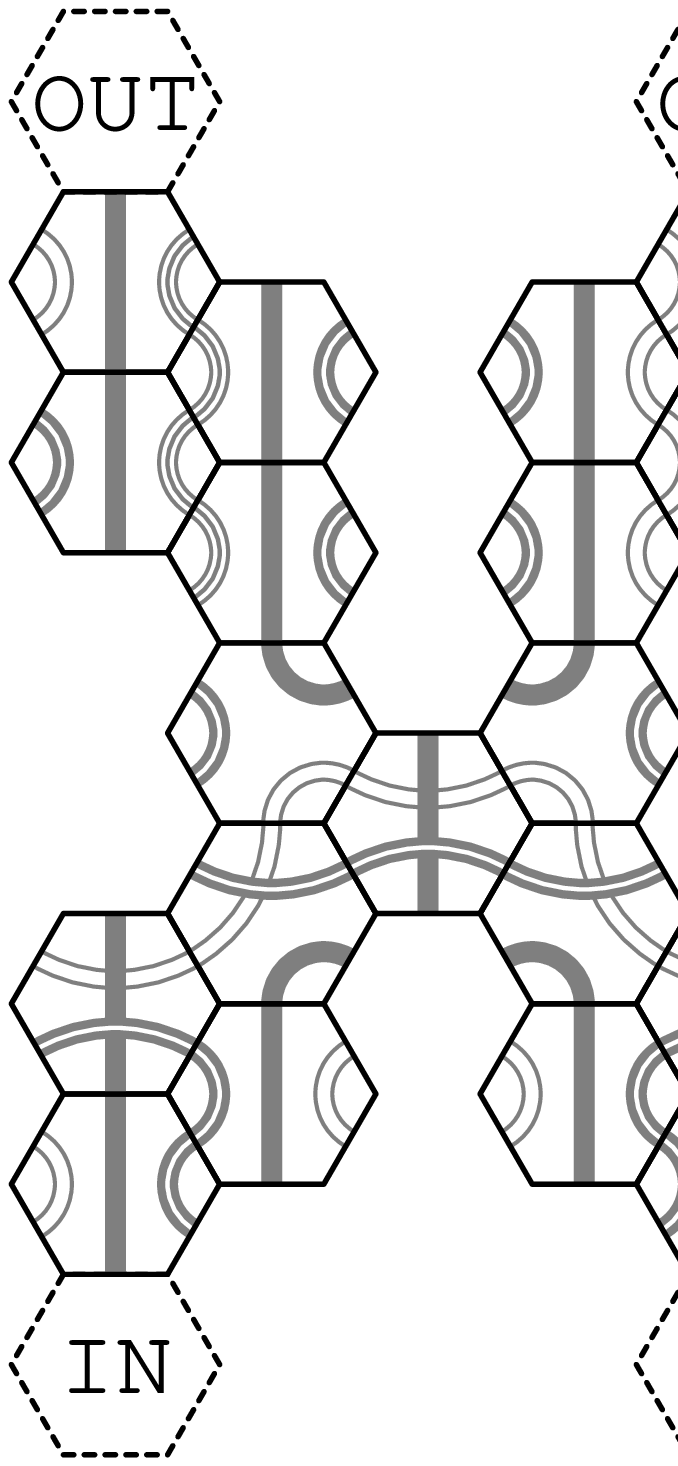}
    \quad
  }
  \subfigure[In: \emph{true, false}]{
    \label{fig:cross-par-4trp-tf}
    \quad
    \includegraphics[height=4.8cm]{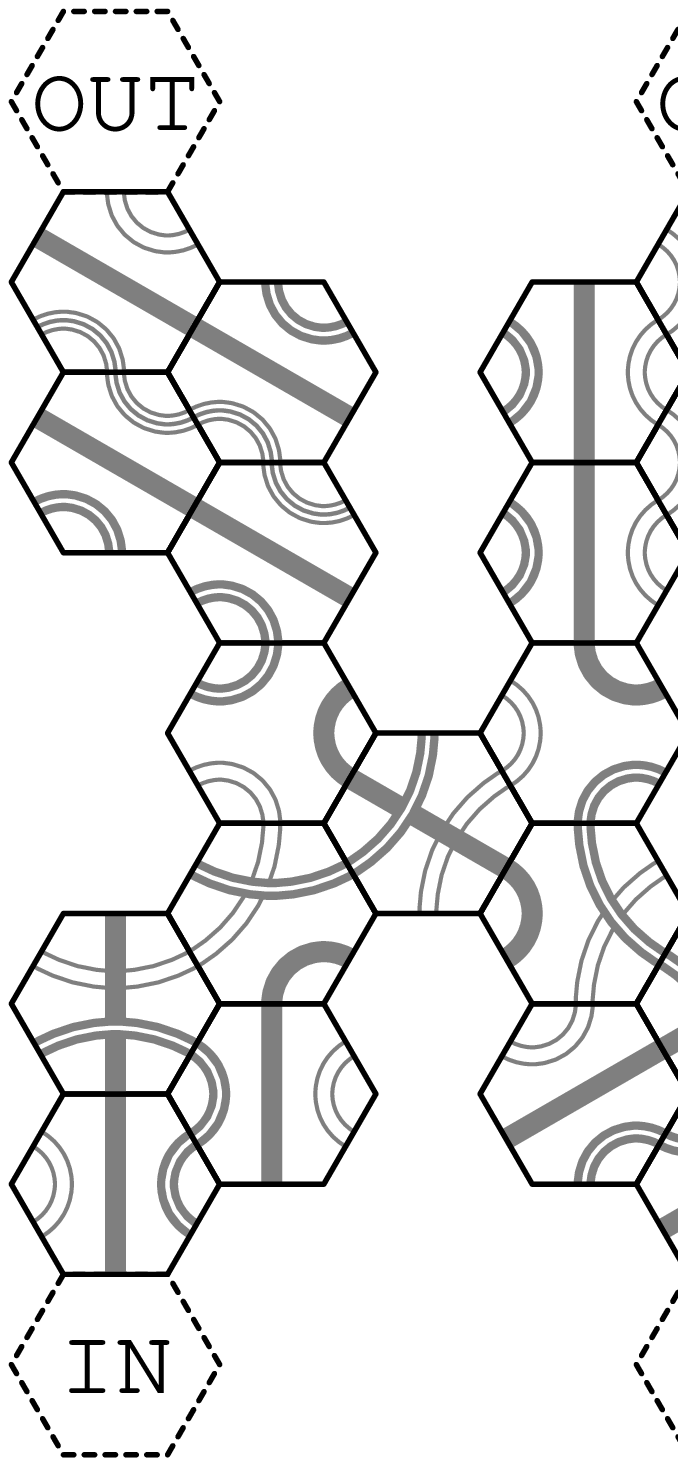}
    \quad
  }  
  \subfigure[In: \emph{false, true}]{
    \label{fig:cross-par-4trp-ft}
    \quad
    \includegraphics[height=4.8cm]{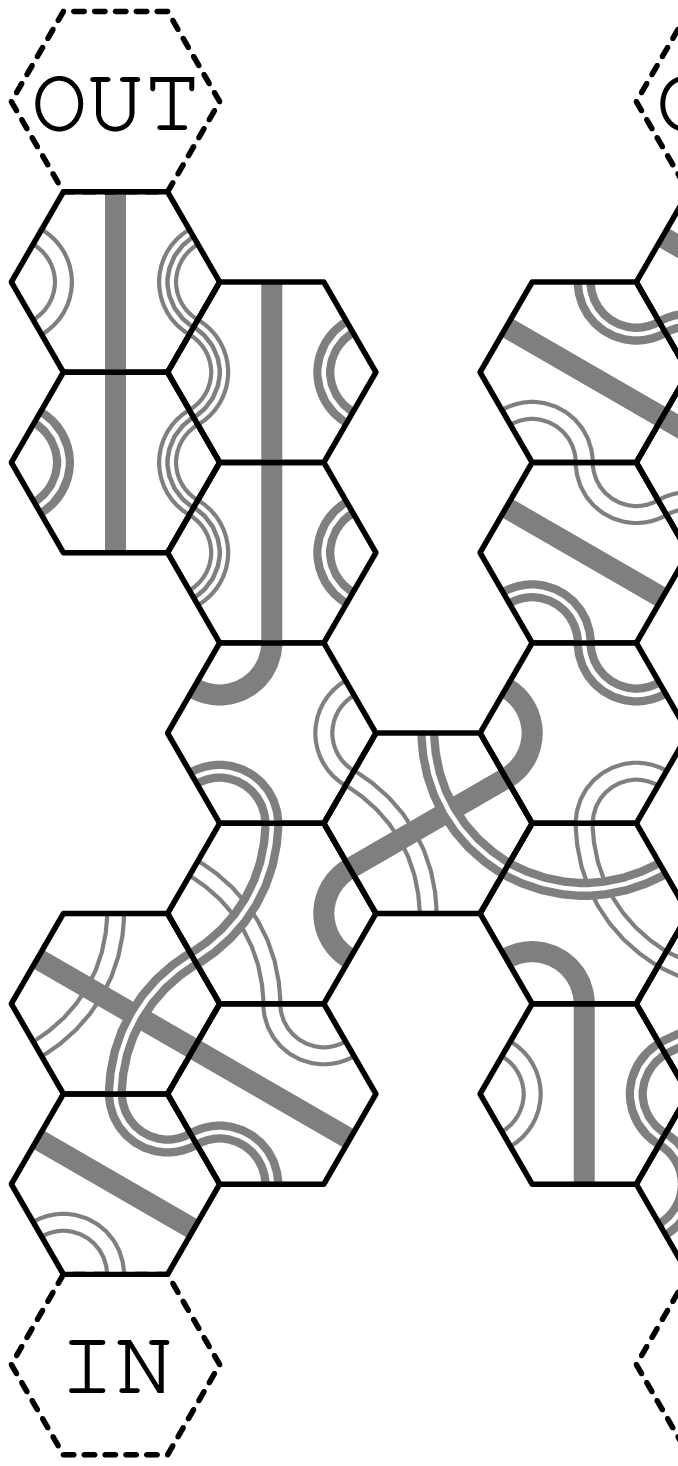}
    \quad
  }
  \subfigure[In: \emph{false, false}]{
    \label{fig:cross-par-4trp-ff}
    \quad
    \includegraphics[height=4.8cm]{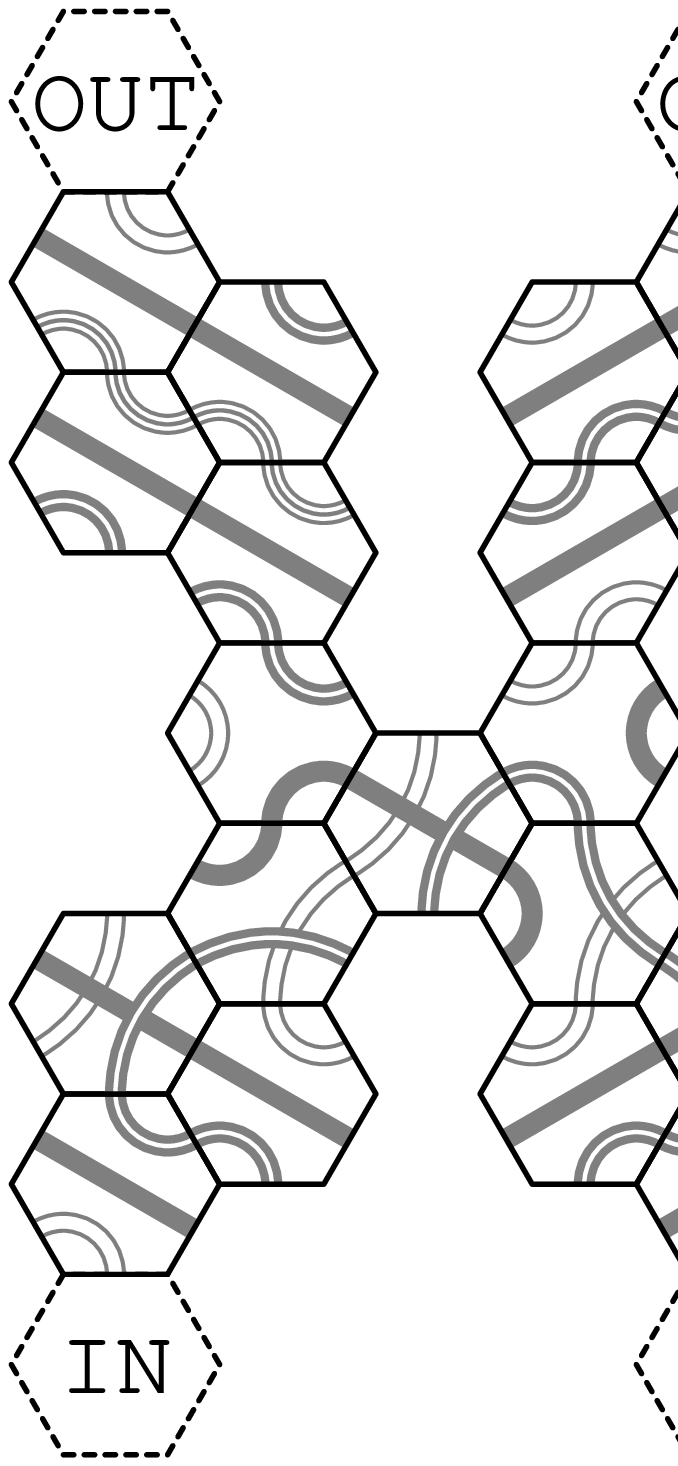}
    \quad
  }
  \subfigure[Scheme]{
    \label{fig:cross-par-4trp-s}
    \quad 
    \includegraphics[height=4.8cm]{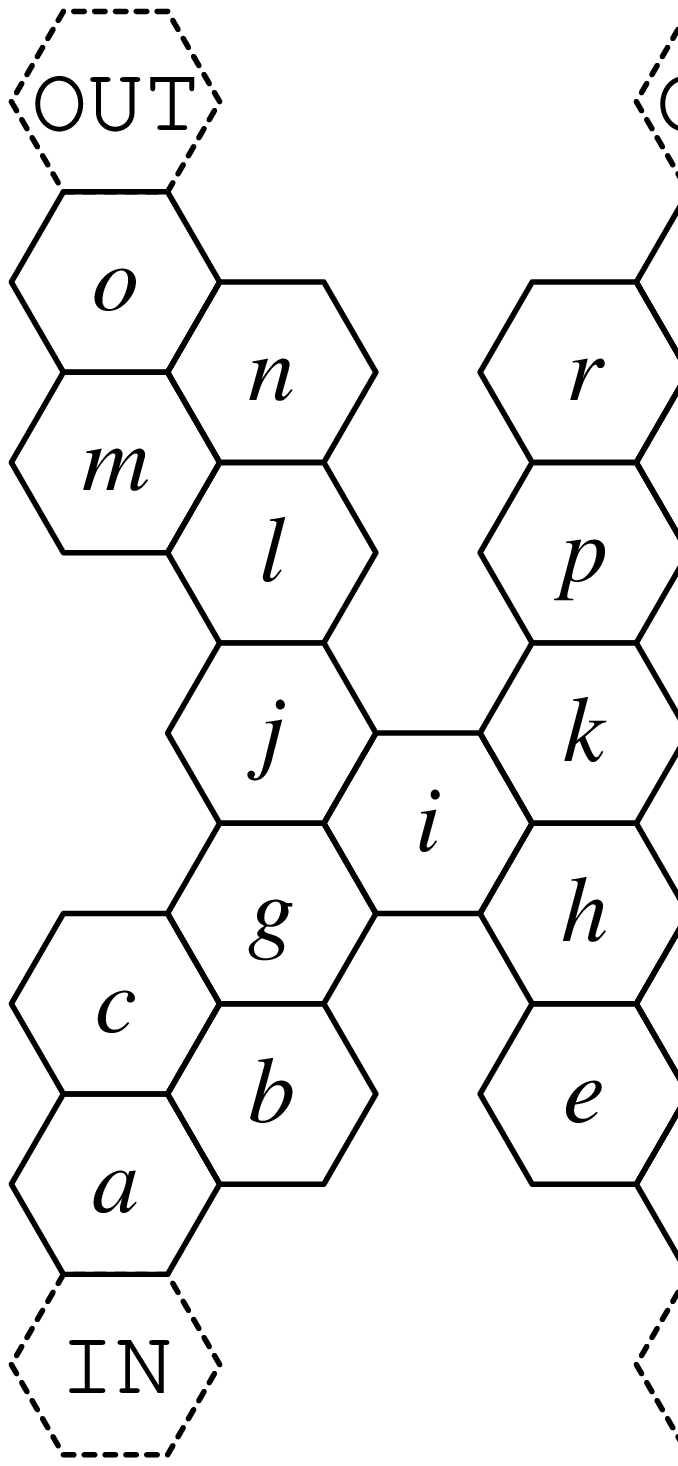}
    \quad 
  }  
  \caption{Subpuzzle CROSS}
  \label{fig:cross-par-4trp}
\end{figure}

To realize wire crossings, Holzer and Holzer~\cite{hol-hol:j:tantrix}
used a circuit construction consisting of AND and NOT gates that is
based on Goldschlager's
procedure~\cite{gol:j:monotone-planar-circuits} and McColl's
``cross-over'' gadget~\cite{mcc:j:planar-crossovers}.  This approach
was also taken in a preliminary version of this
paper~\cite{bau-rot:c:tantrix}.  As mentioned above, however, we here
simplify the construction by simulating wire crossings directly.  To
this end, we introduce a new Tantrix\texttrademark\ subpuzzle called
CROSS, which is presented in Figure~\ref{fig:cross-par-4trp}.  This
subpuzzle has two inputs and two outputs, where the left output will
be the same as the right input and vice versa.

Just as all our modified subpuzzles, our novel CROSS subpuzzle has
unique solutions for each combination of possible input colors.  To
analyze this subpuzzle, we subdivide it into three different
parts. The lower part consists of tiles $a$ through $k$, and the upper
part of tiles $l$ through $t$.

Let us consider the upper part of the CROSS subpuzzle first.  Tiles
$l$ through $o$ form the left output. Since tile $j$ does not contain
\emph{green}, the input color to this part will be either \emph{blue}
or \emph{yellow}.  If the input color is \emph{blue}, all tiles must
have vertical \emph{blue} lines, and since tile $o$ does not contain
\emph{yellow} lines, the orientation of all tiles is fixed with
\emph{green} at the joint edges of, respectively, tiles $o$ and~$n$,
tiles $n$ and~$m$, and tiles $m$ and~$l$. The case of \emph{yellow}
being the input color for this part can be handled similarly and
yields the output color \emph{red}.  The right output consists of
tiles $p$ through~$t$.  Here, the input colors \emph{red},
\emph{blue}, and \emph{yellow} are possible, where \emph{yellow} 
and \emph{red} both lead to output color \emph{red}.  Since tile 
$s$ contains no \emph{red} lines, we obtain unique solutions for all 
possible combinations of input colors, by a similar argument as for 
the left output.

We now turn to the more complicated lower part of this subpuzzle.
Tiles $a$, $b$, and $c$ move the left input color to the joint edge of
tiles $b$ and~$g$.  Regarding tiles~$a$, $b$, and~$c$, there are two
possible solutions for each input color.  The possible color sequences
for the edges of tiles $c$ and $b$ joint with tile $g$ are ${\tt by}$
and ${\tt br}$ for input color \emph{blue}, and are ${\tt ry}$ and
${\tt rb}$ for input color \emph{red}.  Since $g$ contains exactly one
of these color sequences for each input color, the orientation of
tiles~$a$, $b$, and $c$ is fixed, and the input color is passed on to
the joint edge of tiles $b$ and~$g$.  The same argument applies to the
right input, which consists of tiles~$d$, $e$, $f$, and~$h$, since
they are mirror-symmetrical.  If both inputs are \emph{blue}, the
orientations of tiles $g$ and $h$ is fixed by tiles~$b$, $c$, $e$,
and~$f$, respectively.  Both $g$ and $h$ will have \emph{yellow} at
their edges joint with tile~$i$, which leads to \emph{red} at the
edges of tile $i$ joint with tiles $j$ and~$k$, respectively, and the
orientation of all tiles in the lower part is fixed.

The connections between the lower and the upper part are the joint
edges of tiles $j$ and $l$ and of tiles $k$ and~$p$, respectively.
These edges are both \emph{blue} if both input colors are \emph{blue}.
Now consider the case that both input colors are \emph{red}.  The
possible colors for the joint edges of tiles $g$ and $h$ with tile $i$
are \emph{red} and \emph{blue}.  Clearly, it is not possible that they
both are \emph{blue}. Furthermore, it is not possible that the joint
edge of tiles $g$ and $i$ is \emph{blue}, since in this case the joint
edge of tiles $j$ and $l$ would have to be \emph{red}, which is not
possible because $l$ does not contain \emph{red}.  For the same reason
it is not possible that the joint edges of tiles $g$ and $h$ with $i$
are both \emph{red}, so the only possible solution is that the joint
edge of tiles $g$ and $i$ is \emph{red} and that the joint edge of
tiles $h$ and $i$ is \emph{blue}.  This also uniquely determines the
orientation of tiles $j$ and $k$ with \emph{yellow} at the joint edge
of tiles $j$ and $l$ and \emph{red} at the joint edge of tiles $k$
and~$p$.  The cases that one input is \emph{red} and the other one is
\emph{blue} can be handled by similar arguments to show that we then
have unique solutions as well.\\

% DB: Reicht dass oder soll ich die anderen beiden Faelle auch
% noch ausfuehrlich begruenden?
% Also dass auch dann die richtige Loesung rauskommt? 
%%% JR: Ich denke, das reicht so.

\subsection{Gate Subpuzzles}
\label{sec:gate}

The gates of the boolean circuit are simulated by the corresponding
AND and NOT subpuzzles.

\begin{figure}[h]
  \centering
  \subfigure[In: \emph{true}]{
    \label{fig:not-4trp-t}
    \quad
    \includegraphics[height=3.6cm]{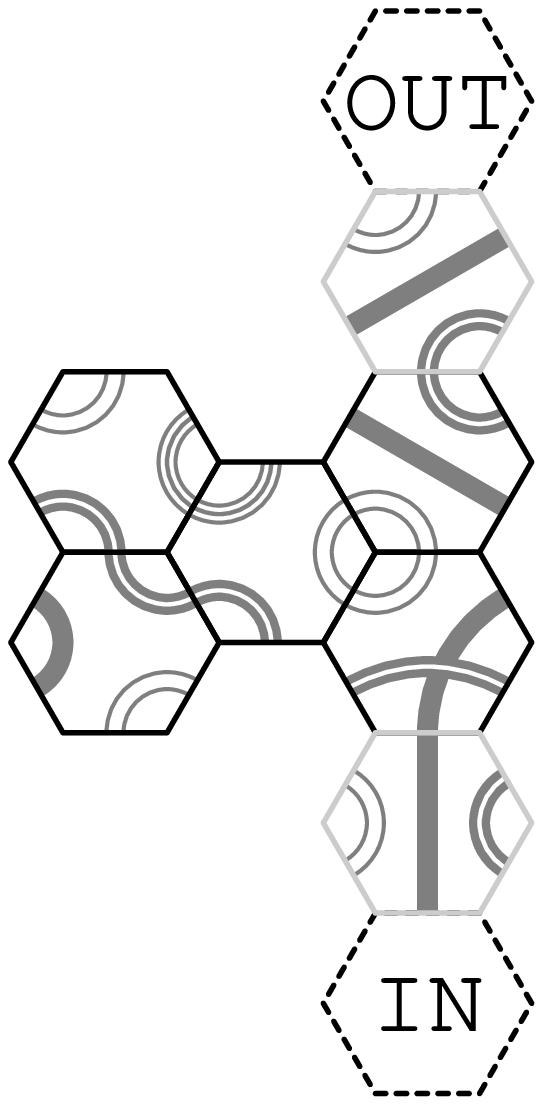}
    \quad
  }
  \subfigure[In: \emph{false}]{
    \label{fig:not-4trp-f}
    \quad
    \includegraphics[height=3.6cm]{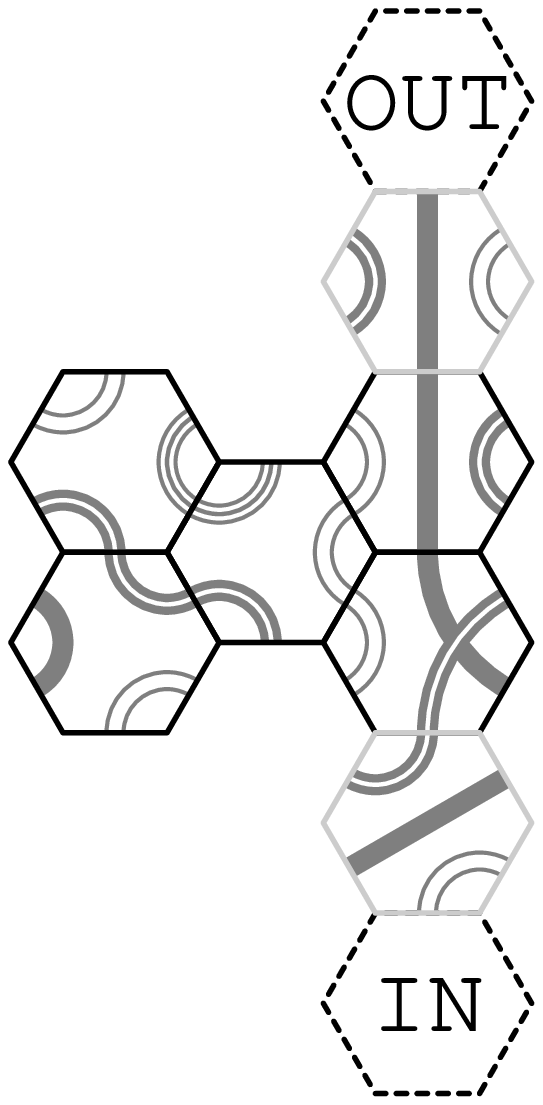}
    \quad
  }   
  \subfigure[Scheme]{
    \label{fig:not-4trp-s}
    \quad 
    \includegraphics[height=3.6cm]{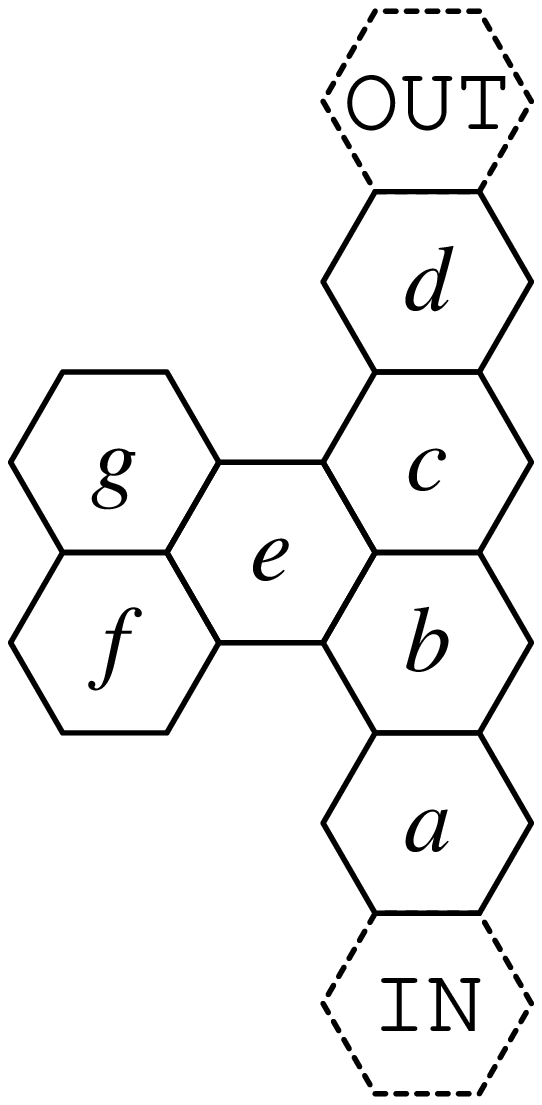}
    \quad 
  }   
  \caption{Original subpuzzle NOT, see~\cite{hol-hol:j:tantrix}}
  \label{fig:not-4trp}
\end{figure}

\begin{figure}[h!]
  \centering
  \subfigure[In: \emph{true}]{
    \label{fig:not-par-4trp-t}
    \quad 
    \includegraphics[height=3.6cm]{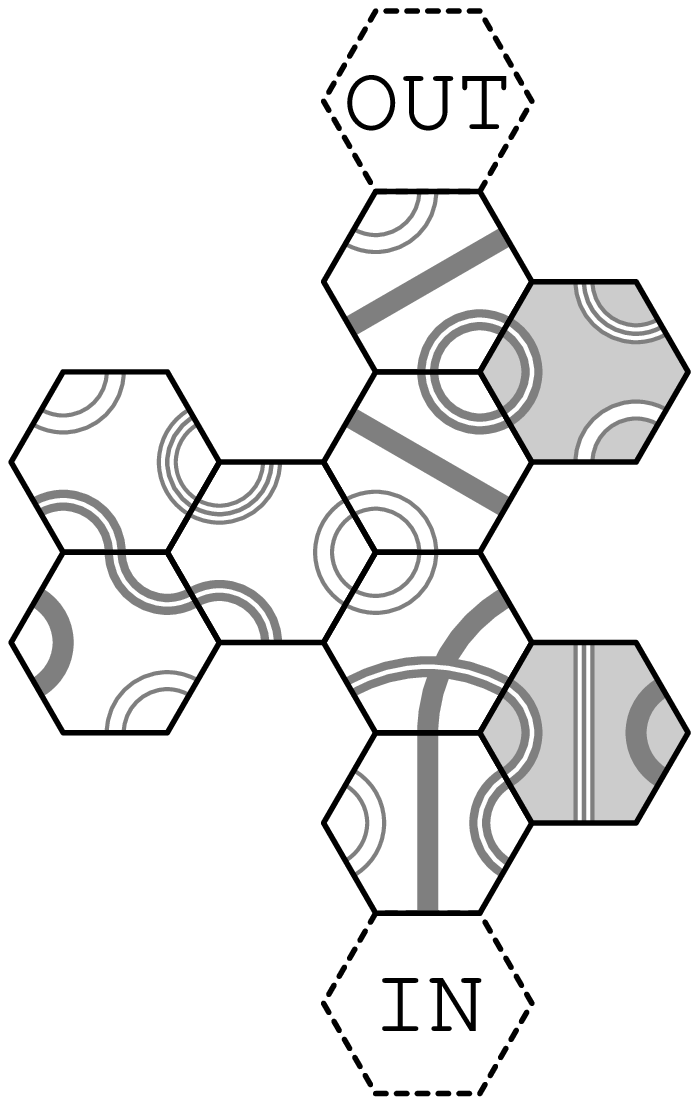}
    \quad 
  }
  \subfigure[In: \emph{false}]{
    \label{fig:not-par-4trp-f}
    \quad 
    \includegraphics[height=3.6cm]{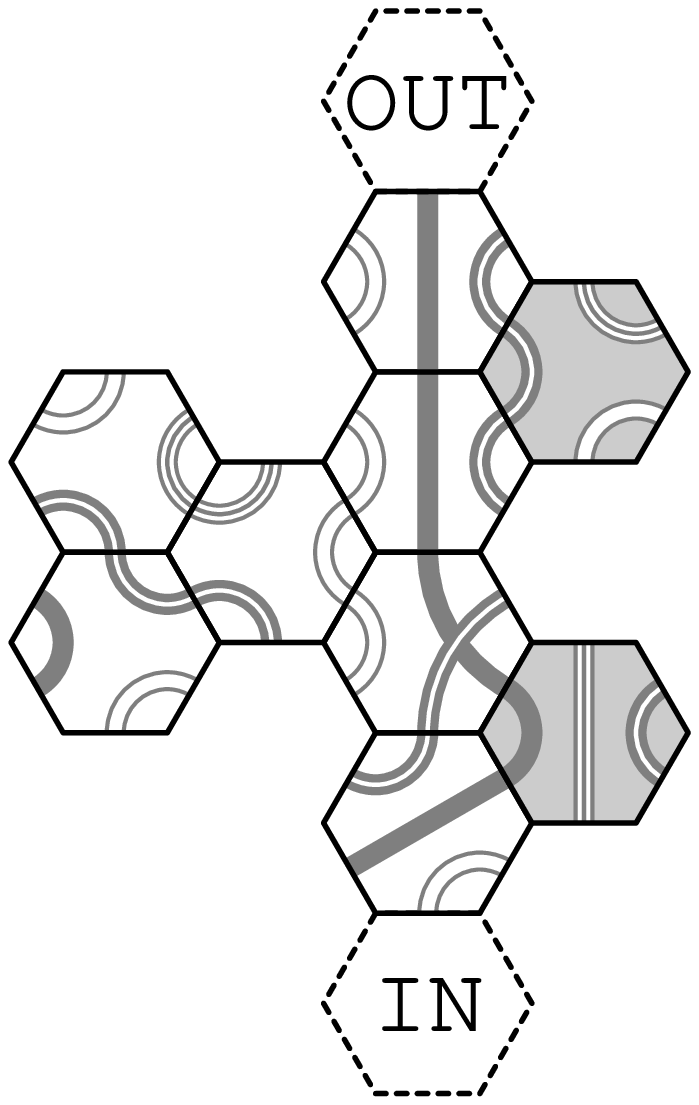}
    \quad 
  }
  \subfigure[Scheme]{
    \label{fig:not-par-4trp-s}
    \quad 
    \includegraphics[height=3.6cm]{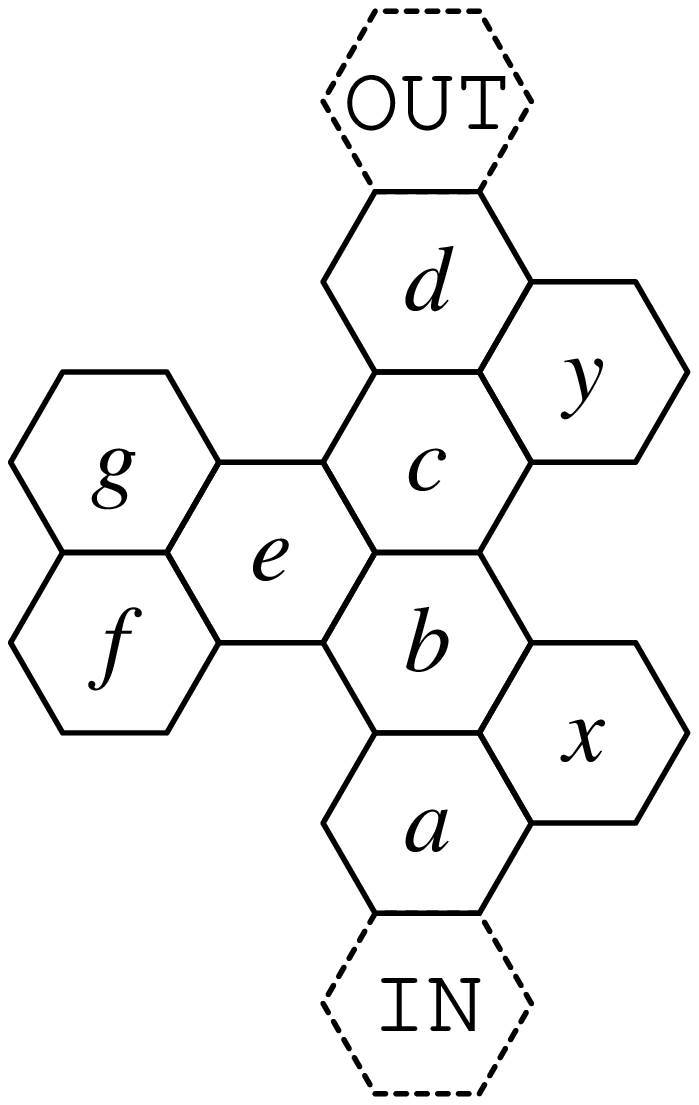}
    \quad 
  }   
  \caption{Modified subpuzzle NOT}
  \label{fig:not-par-4trp}
\end{figure}

The original version of the NOT subpuzzle
from~\cite{hol-hol:j:tantrix} is presented in
Figure~\ref{fig:not-4trp}, and our new version is shown in
Figure~\ref{fig:not-par-4trp}.  The purpose of this subpuzzle is to
``negate'' the input color, i.e., if the input color is \emph{red}
then the output color will be \emph{blue}, and vice versa.

In the original subpuzzle, there is only one possible solution to the
three Ronds $e$, $f$, and~$g$: Since tiles $c$, $b$, and $f$ do not
contain \emph{green}, the joint edge of tiles $e$ and $g$ must be
\emph{green}, which forces the joint edge of tiles $g$ and $f$ to be
\emph{yellow}.  So the orientation of tiles $e$, $f$, and $g$ is
fixed, with \emph{red} at the edges joint with tiles $b$ and~$c$,
respectively.  There are only two possible orientations left for the
tiles $b$ and~$c$, one for input color \emph{blue} and one for input
color \emph{red}.  The only tiles still having more than one possible
orientation are $a$ and~$d$.  They will be fixed by inserting tiles
$x$ and~$y$.  The possible color sequences for the edges of tiles $a$
and $b$ joint with tile $x$ are ${\tt yy}$ or ${\tt ry}$
if the input color is \emph{blue}, and they are ${\tt bb}$ or ${\tt yb}$
if the input color is \emph{red}.  Tile $x$, however, contains only the 
sequences ${\tt yy}$ and ${\tt bb}$, so
the orientation is fixed.  Note that tiles $c$ and $d$ behave just
like a WIRE subpuzzle and since the orientation of tile $c$ is fixed
with \emph{red} at the edge joint with tile~$e$, we insert a Rond at 
position $y$ containing the color sequence ${\tt yy}$ but none of the 
sequences ${\tt yb}$ and ${\tt yr}$.  With these two new tiles, unique 
solutions are enforced for each input color. \\

% Figure~\ref{fig:not-par-4trp} presents the modified NOT subpuzzle, which
% negates the input value by flipping the colors blue and red.  In the
% original NOT subpuzzle from~\cite{hol-hol:j:tantrix} (see
% Figure~\ref{fig:not-4trp} in the appendix), there is only one
% possible orientation for the tiles $e$, $f$, and $g$, since the tiles
% $c$, $b$, and $f$ do not contain green.  Thus tiles $c$ and $b$ must
% have red at the edge adjacent to tile~$e$.  It follows that for each
% input color only two orientations are possible for tiles $a$ and~$d$.
% Inserting a \emph{Brid} in the colors blue, yellow, and green at
% position $x$ uniquely determines the orientation of tile~$a$.  Since
% $x$ does not contain red, we have that tile $a$ is forced to choose
% yellow at the edge adjacent to $x$ if the input color was blue.  On
% the other hand, if the input color was red, $a$ has a choice between
% blue and green for this color because $x$ has blue at the edge joint
% with~$b$.  However, since $a$ doesn't contain green, this uniquely determines
% the orientation of~$a$.  The orientation of tile $d$ can be made
% unique by inserting a \emph{Rond} in the colors yellow, red, and green
% at position~$y$.  For both input colors, $c$ has yellow at the edge
% joint with~$y$, so $d$ and $y$ can share either yellow or green.
% Since tile $d$ contains no green, its orientation is uniquely
% determined.
% % in both cases.
% Thus, for both input colors, a valid solution of the NOT subpuzzle is
% uniquely determined.

\begin{figure}[h]
  \centering
  \subfigure[In: \emph{true, true}]{
    \label{fig:and-4trp-tt}
    \quad
    \includegraphics[height=4.8cm]{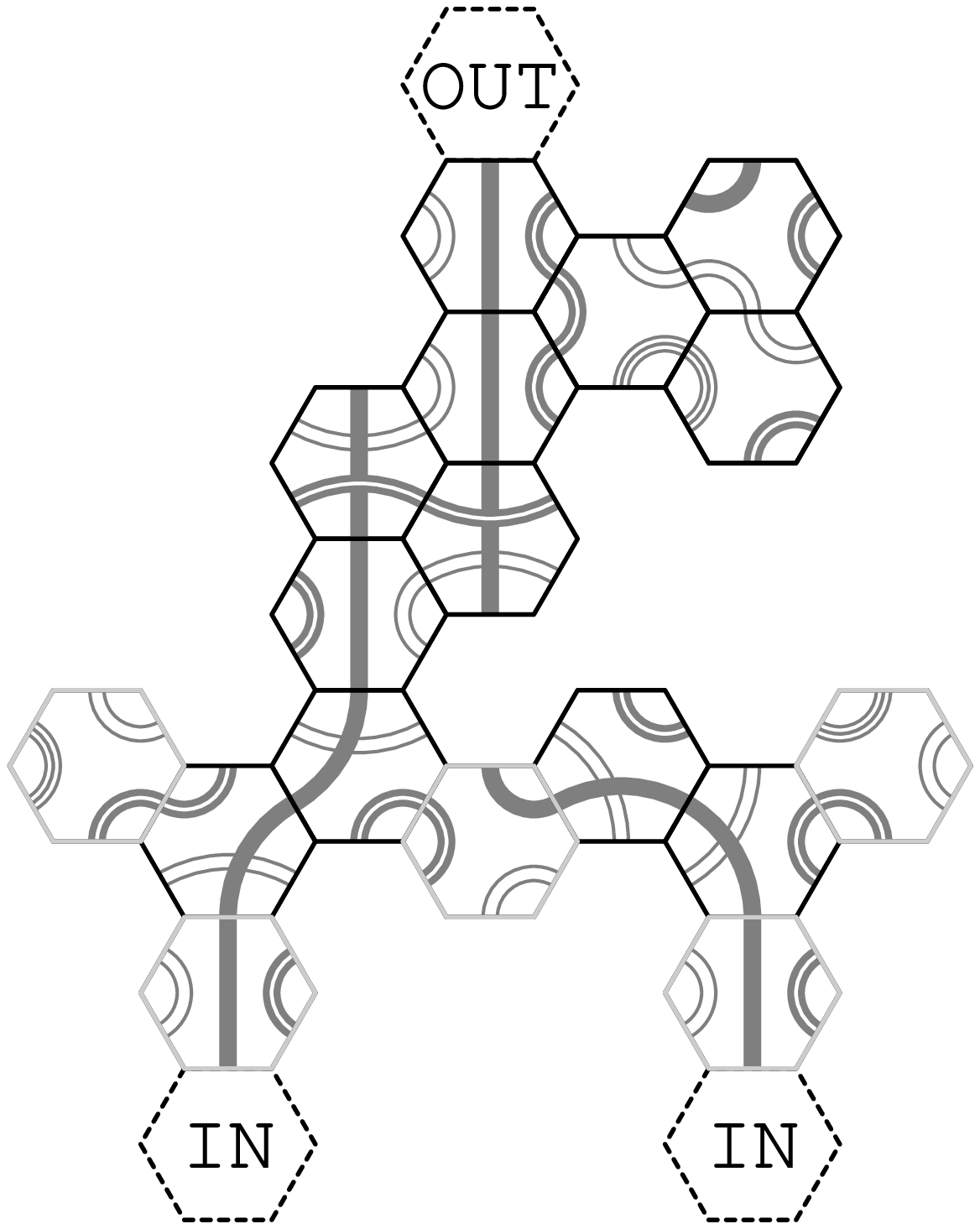}
    \quad
  }
  \subfigure[In: \emph{true, false}]{
    \label{fig:and-4trp-tf}
    \quad
    \includegraphics[height=4.8cm]{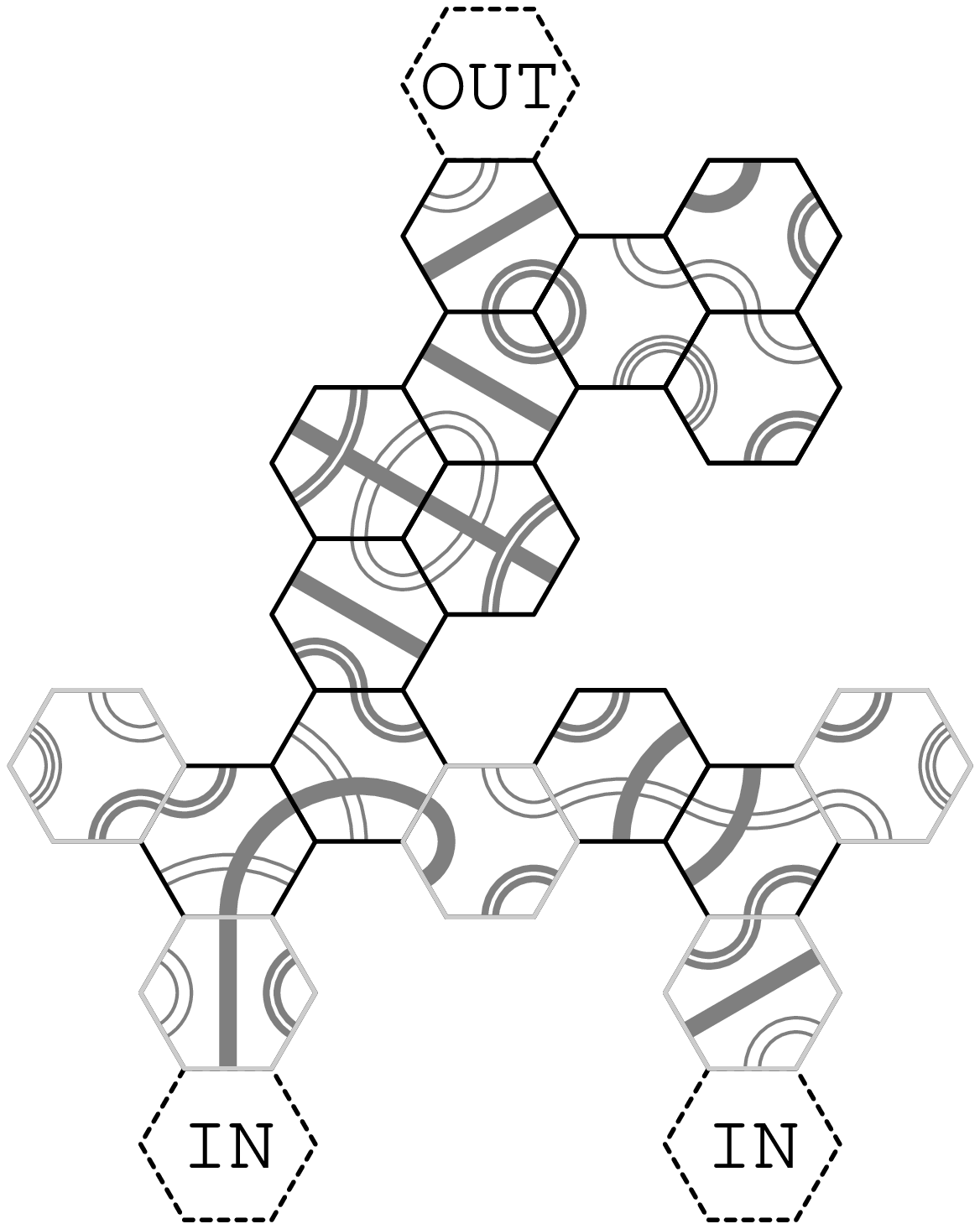}
  }  
  \subfigure[In: \emph{false, true}]{
    \label{fig:and-4trp-ft}
    \quad
    \includegraphics[height=4.8cm]{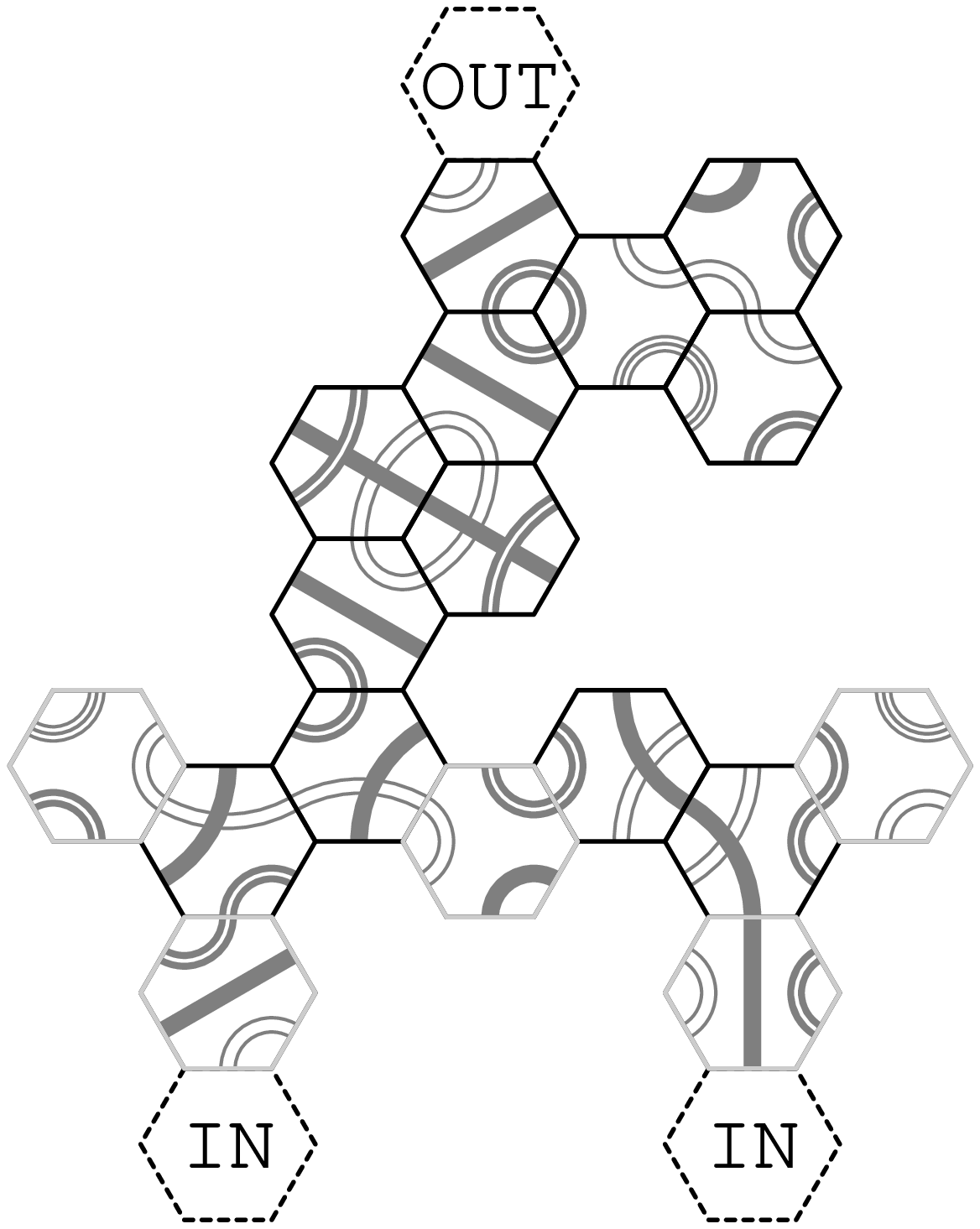}
    \quad
  }
  \subfigure[In: \emph{false, false}]{
    \label{fig:and-4trp-ff}
    \quad
    \includegraphics[height=4.8cm]{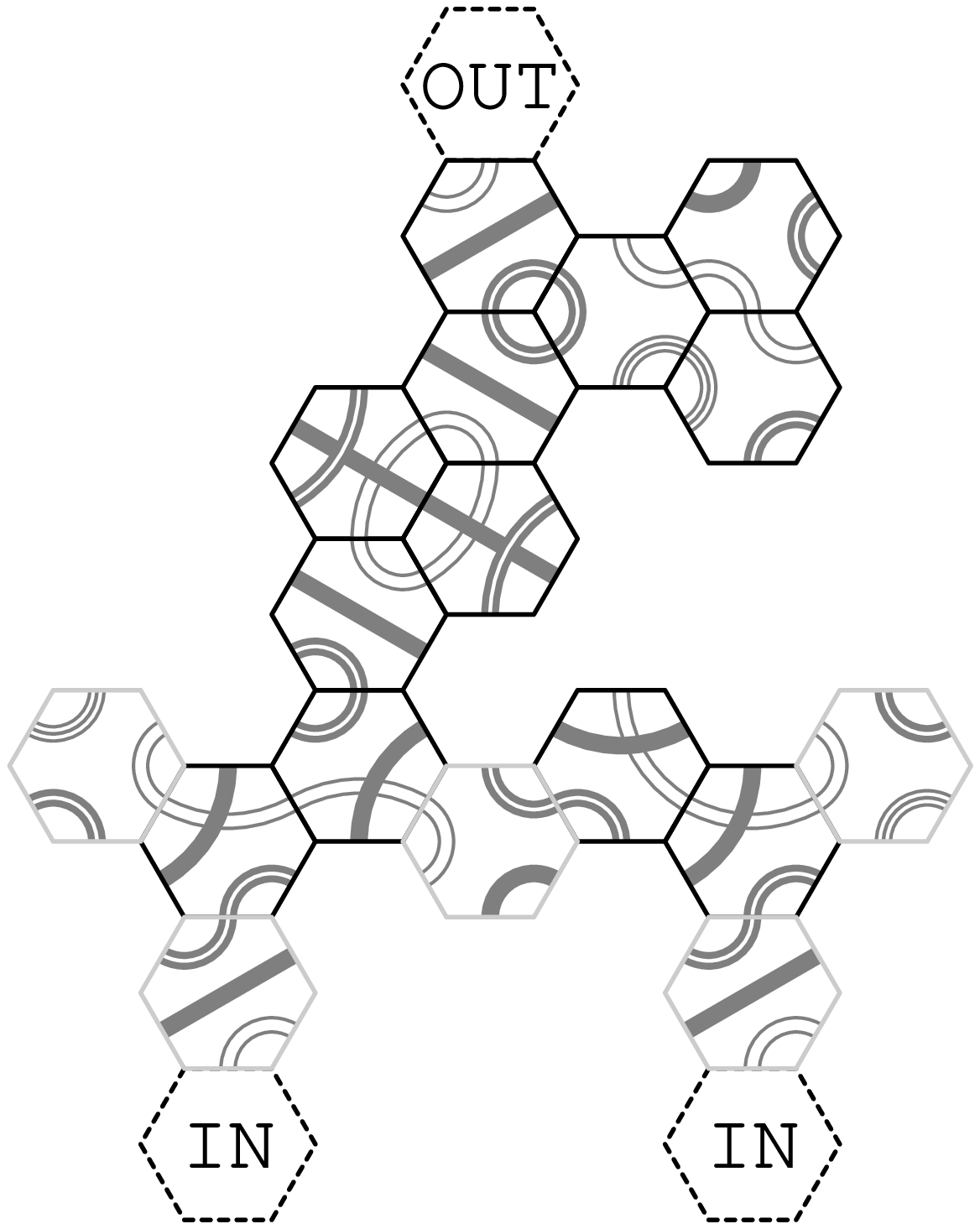}
    \quad
  }
  \subfigure[Scheme]{
    \label{fig:and-4trp-s}
    \quad 
    \includegraphics[height=4.8cm]{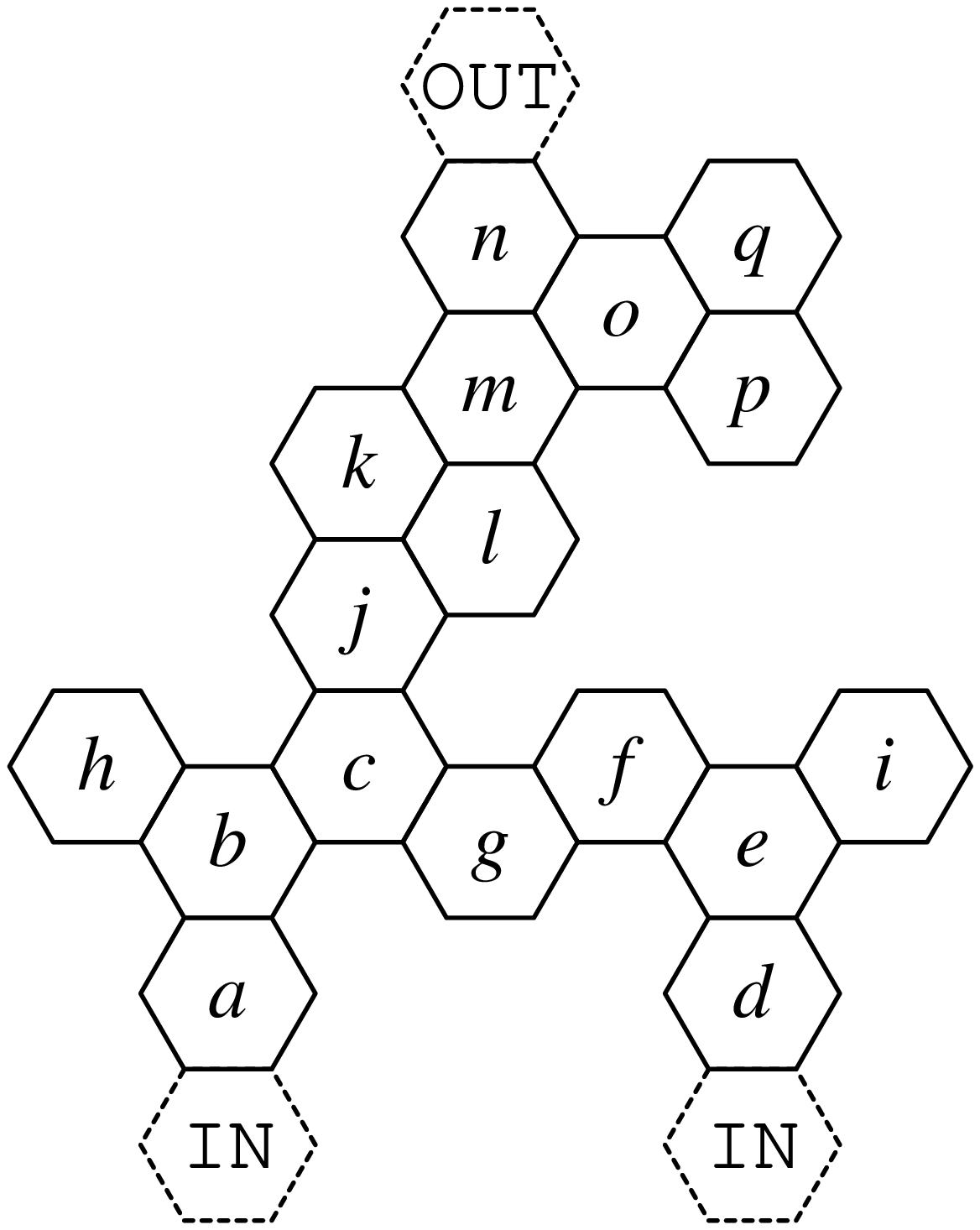}
    \quad 
  }   
  \caption{Original subpuzzle AND, see~\cite{hol-hol:j:tantrix}}
  \label{fig:and-4trp}
\end{figure}

\begin{figure}[t!]
  \centering

  \subfigure[In: \emph{true, true}]{
    \label{fig:and-par-4trp-tt}
    \quad
    \includegraphics[height=4.8cm]{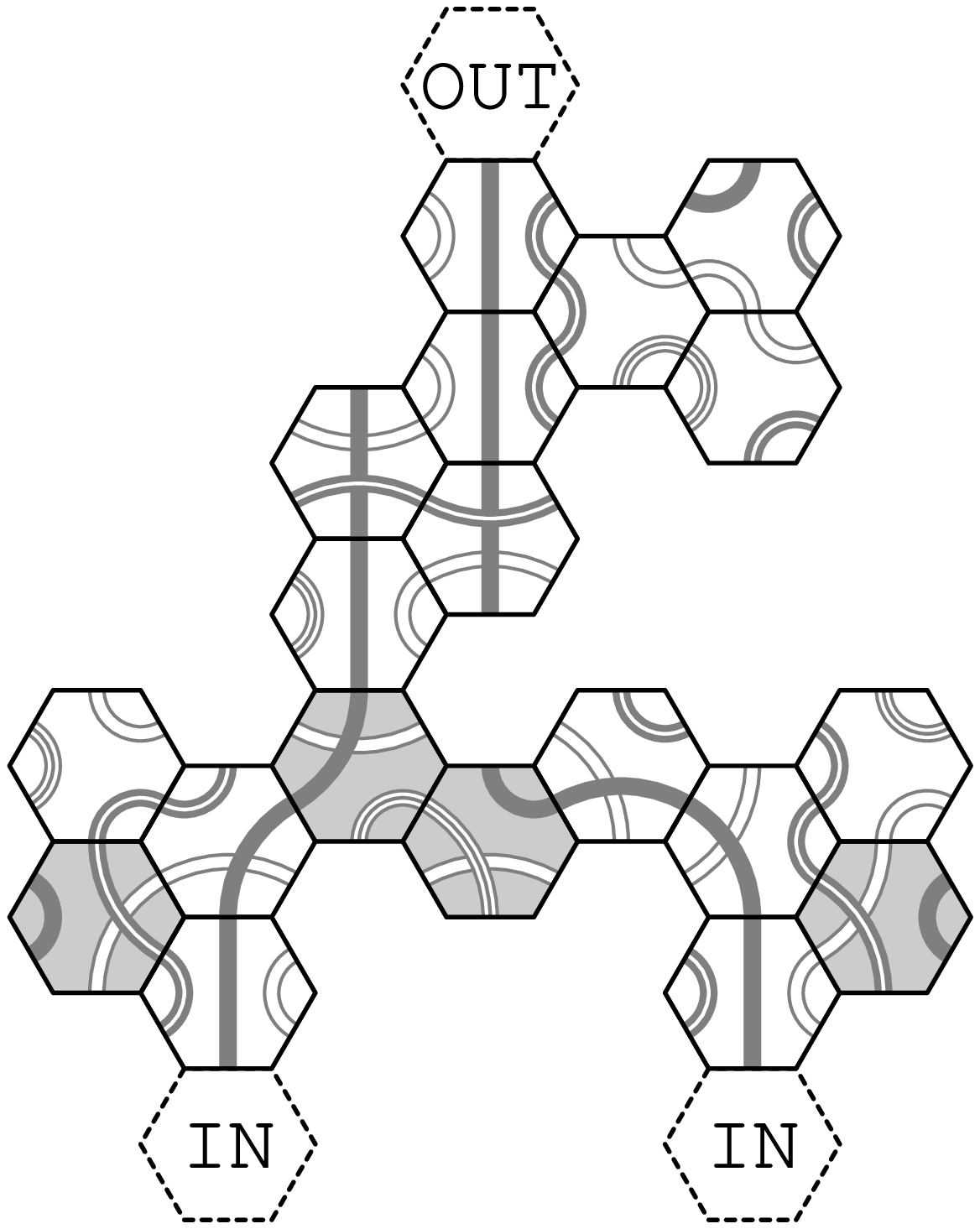}
    \quad 
  }
  \subfigure[In: \emph{true, false}]{
    \label{fig:and-par-4trp-tf}
    \quad 
    \includegraphics[height=4.8cm]{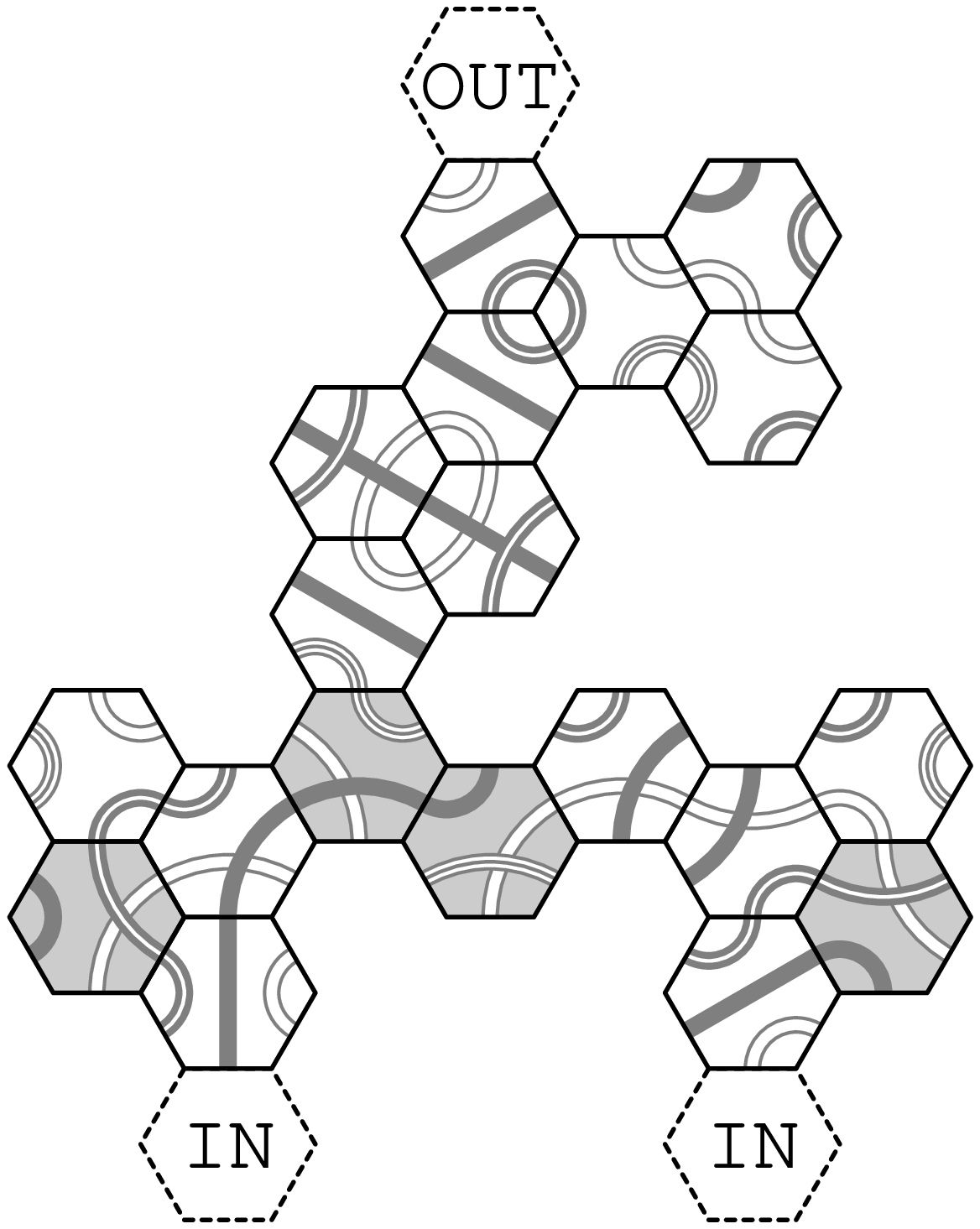}
    \quad 
  }  
  \subfigure[In: \emph{false, true}]{
    \label{fig:and-par-4trp-ft}
    \quad 
    \includegraphics[height=4.8cm]{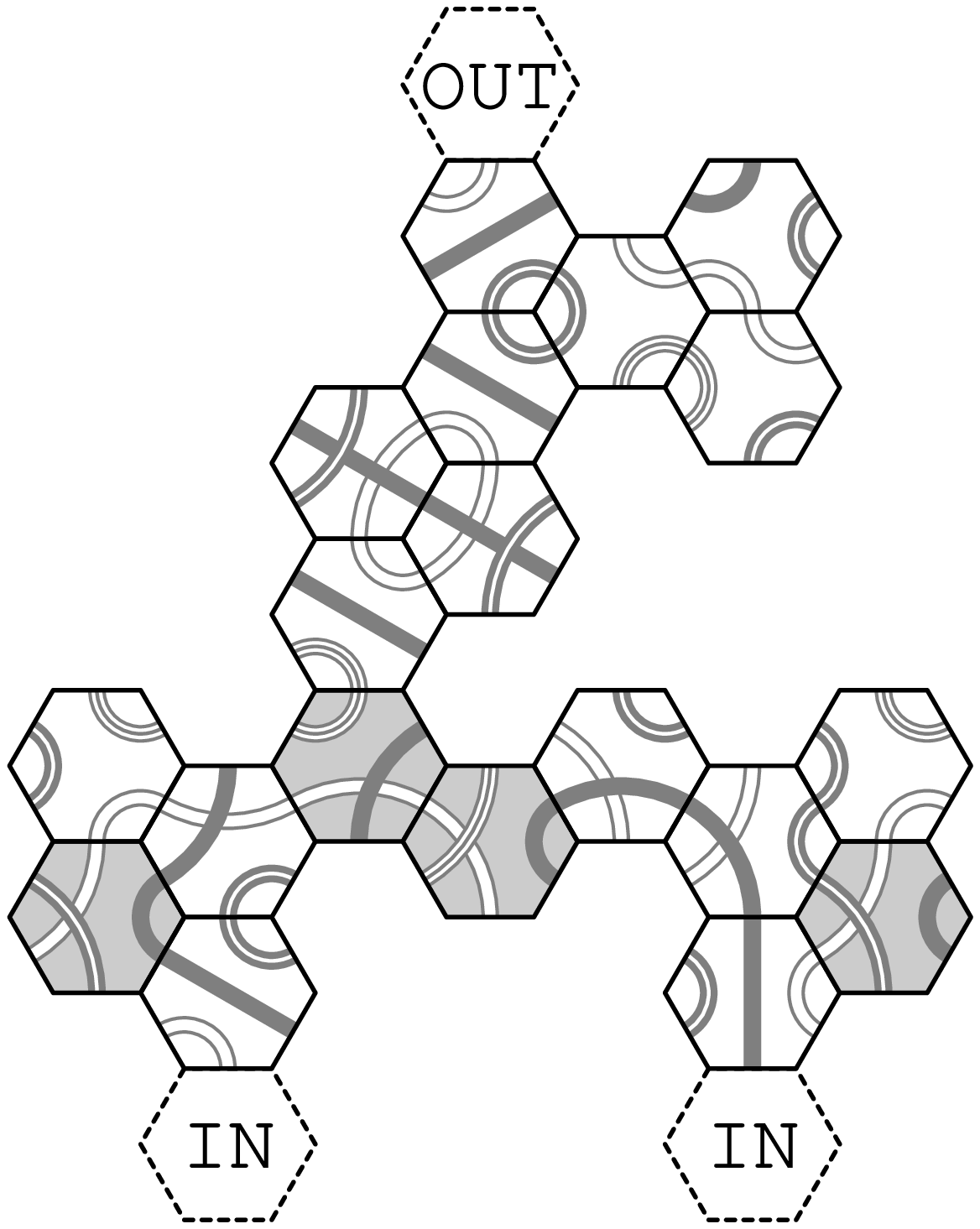}
  }
  \subfigure[In: \emph{false, false}]{
    \label{fig:and-par-4trp-ff}
    \quad 
    \includegraphics[height=4.8cm]{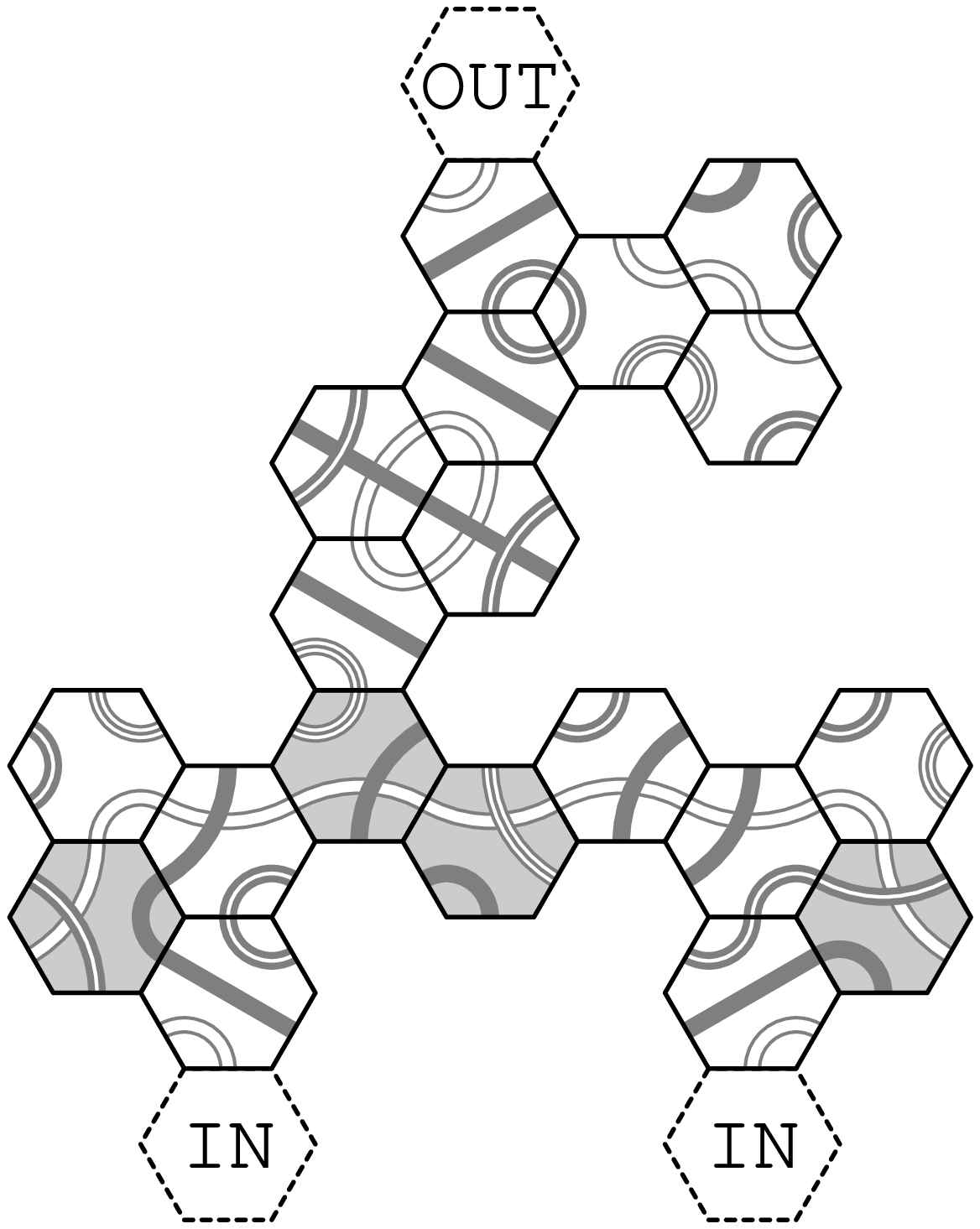}
    \quad 
  }
  \subfigure[Scheme]{
    \label{fig:and-par-4trp-s}
    \quad 
    \includegraphics[height=4.8cm]{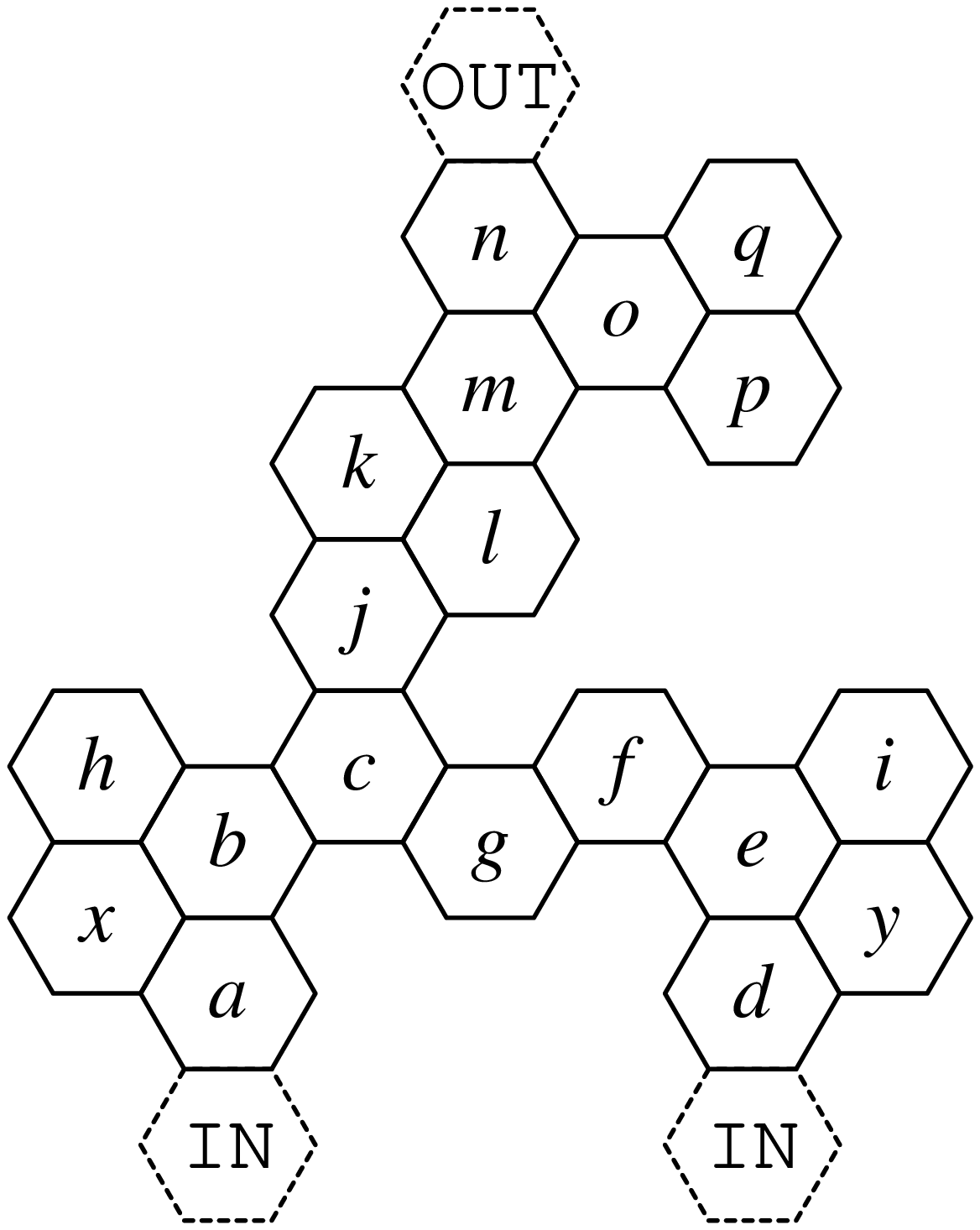}
    \quad 
  }  
  \caption{Modified subpuzzle AND}
  \label{fig:and-par-4trp}
\end{figure}

The somewhat more complicated AND subpuzzle is shown in
Figure~\ref{fig:and-4trp} in its original version
from~\cite{hol-hol:j:tantrix}, while Figure~\ref{fig:and-par-4trp}
presents our modified version.

This subpuzzle can again be subdivided
into two different parts, an upper part and a lower part that are
connected at the joint edge of tiles $c$ and~$j$.

First, we consider the upper part of the modified AND subpuzzle.  We
show that its output color is \emph{red} if the joint edge of tiles
$j$ and $c$ is \emph{yellow}, and that its output color is \emph{blue}
if this edge is \emph{blue}.  Note that, just as for the NOT
subpuzzle, the Ronds $o$, $p$, and $q$ have only one possible
orientation, thus forcing the edges of tiles $n$ and $m$ joint with
tile $o$ to be \emph{yellow}.  If the input to this part of the subpuzzle is
\emph{blue} then tiles $j$ and $k$ must have a vertical \emph{blue}
line, and since the edges of tile $l$ joint with $j$ and $k$ cannot be
\emph{blue}, tiles $l$, $m$, and $n$ must have vertical \emph{blue}
lines, too.  Since the edges of tiles $n$ and $m$ joint with tile $o$
are \emph{yellow}, the orientation of all other tiles is uniquely
determined in the upper part.  The case of \emph{yellow} being the
input color for the upper part can be handeld similarly: Without any
modification, we have unique solutions for the upper part of the
subpuzzle.

Now we consider the lower part of this subpuzzle.  For each
combination of input colors, tiles $a$ and~$d$ (which are adjacent to
the input tiles of the AND subpuzzle) and tiles $h$ and $i$ (having
only one connecting edge to the rest of the original AND subpuzzle in
Figure~\ref{fig:and-4trp}) each have two possible orientations.
Examining all possible color sequences for each combination of
possible input colors, tiles $x$ and $y$ can be determined as shown in
Figure~\ref{fig:and-par-4trp} to fix the orientation of tiles $a$ and
$h$ and of tiles $d$ and~$i$, respectively.

Then, the input color is passed on to the joint edge of tiles $b$
and~$c$, and to the joint edge of tiles $e$ and $f$.  Tile $g$ in the
center of the subpuzzle has again two possible orientations for each
combination of input colors.  Note that our modifications made so
far---namely, inserting new tiles into the subpuzzles---do not work
here, as it is not possible to insert a new tile in the neighborhood
of tile~$g$, and thus we have to replace it.  As we have seen in the
analysis of the upper part, the color \emph{yellow} of tile $j$ is
only used for the edge joint with tile~$c$.  We will replace the color
\emph{yellow} of both tiles with the color \emph{green}.  This is
possible, because the joint edge of tiles $c$ and $b$ will never be
\emph{yellow}, and tile $g$ will be replaced by a new one.  To do
this, we consider all possible colors at the edges of tiles $c$ and
$f$ joint with tile~$g$, with the restriction that the joint edge of
tiles $c$ and $j$ must be either \emph{blue} or \emph{green}.

If the right input color is \emph{blue} then the joint edge of tiles
$f$ and $g$ can be either \emph{blue} or \emph{yellow}, and if the
right input color is \emph{red} then this edge can be either
\emph{red} or \emph{yellow}.  For the left input, the joint edge of
tiles $c$ and $g$ must be \emph{red} if the input is \emph{red},
because then the joint edge of tiles $c$ and $j$ cannot be
\emph{red}.

If both inputs are \emph{blue} then the output color of the lower part
must be \emph{blue} as well, and then the joint edge of tiles $c$ and
$g$ is \emph{green}.  If only the left input is \emph{blue}, the output 
color of the lower part must be \emph{green}, and the joint edge of 
tiles $c$ and $g$ is \emph{blue}. This leads to the following 
restrictions for tile $g$:
\begin{enumerate}
\item If the input colors are both \emph{blue} then $g$ must contain
exactly one of the color sequences ${\tt g x b}$ or ${\tt g x y}$,
where ${\tt x}$ stands for an arbitrary color.

\item If the right input color is \emph{red} and the left input color
is \emph{blue}, $g$ must contain either ${\tt b x r}$ or ${\tt
b x y}$.

\item If the right input color is \emph{blue} and and the left input
color is \emph{red} then the possible color sequences for $g$ are
either ${\tt r x b}$ or ${\tt r x y}$.

\item For the last combination of both input colors being \emph{red},
tile $g$ must contain either ${\tt r x r}$ or ${\tt r x y}$.
\end{enumerate}

In our modified AND subpuzzle, we insert a Sint instead of a Rond,
which contains each of the color sequences ${\tt g x b}$, ${\tt b x
r}$, ${\tt r x b}$, and ${\tt r x r}$ exactly once.  This leads to
unique solutions for the lower part and thus for the whole subpuzzle,
for each possible combination of input colors.\\

\subsection{Input and Output Subpuzzles}
\label{sec:input-output}

The variables of the boolean circuit are represented by the subpuzzle
{BOOL}.  Holzer and Holzer~\cite{hol-hol:j:tantrix} showed that this
subpuzzle has only two solutions, one with output color \emph{blue}
and the other one with output color \emph{red}.  This output color
corresponds to the boolean value of the corresponding input variable,
where \emph{blue} stands for the truth value \emph{true}, and
\emph{red} stands for \emph{false}.  The restriction that
\emph{yellow} and \emph{green} are not possible as output colors for
this subpuzzle, ensures that the following subpuzzles will always have
a valid input color, namely \emph{blue} or \emph{red}.\\

% The BOOL subpuzzle represents the input gates of the circuit.  This
% subpuzzle has only two valid solutions, either its output is blue (if
% the corresponding input variable is \emph{true}), or it is red (if
% the corresponding input variable is \emph{false}).  This ensures that
% subsequent subpuzzles can obtain only these two colors as input.

The last subpuzzle needed to simulate a boolean circuit is the
subpuzzle {TEST}. It is placed at the output gate of the circuit, and
its purpose is to verify that the circuit evalutes to true. Holzer and
Holzer~\cite{hol-hol:j:tantrix} mention that this subpuzzle has only
one valid solution with \emph{blue} as the input color.  This ensures
that the output of the whole circuit will be \emph{true} if and only
if the Tantrix\texttrademark\ puzzle constructed has a solution.  Like
the original subpuzzle BOOL, the original TEST subpuzzle already has a
unique solution, so it is not modified.

\begin{figure}[h!]
  \centering
  \subfigure[BOOL Out: \emph{true}]{
    \label{fig:bool-4trp-t}
    \quad
    \quad
    \quad
    \includegraphics[height=1.8cm]{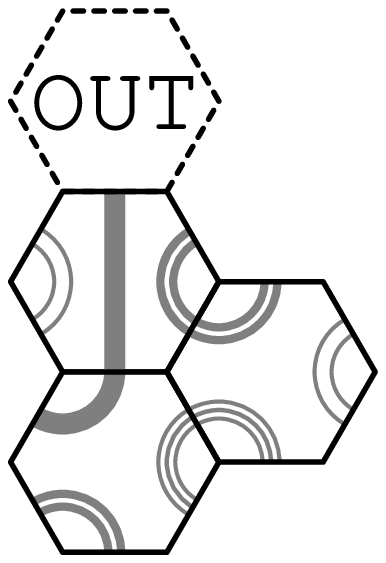}
    \quad
    \quad
    \quad
  }
  \subfigure[BOOL Out: \emph{false}]{
    \label{fig:bool-4trp-f}
    \quad
    \quad
    \quad
    \includegraphics[height=1.8cm]{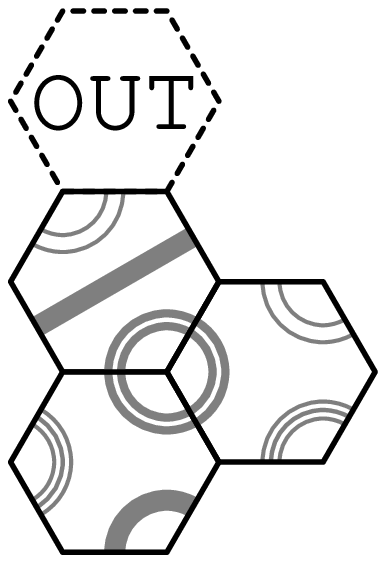}
    \quad
    \quad
    \quad
  }
%   \subfigure[Scheme]{
%     \label{fig:bool-4trp-s}
%     \quad
%     \quad
%     \includegraphics[height=1.8cm]{../Bilder-EPS-sw/bool-4trp-s}
%     \quad
%     \quad
%   }
  \subfigure[TEST]{
    \label{fig:test-4trp-t}
    \quad
    \quad
    \quad
    \includegraphics[height=1.8cm]{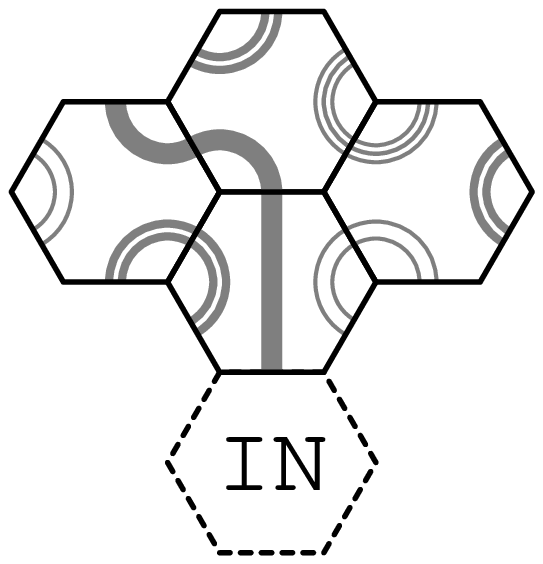}
    \quad
    \quad
    \quad
  }
  \caption{Subpuzzles BOOL and TEST, see~\cite{hol-hol:j:tantrix}}
  \label{fig:bool-test-4trp}
\end{figure}

These two subpuzzles are the only ones from~\cite{hol-hol:j:tantrix}
that are not modified.  For the sake of completeness, they are
presented in 
%Figures~\ref{fig:bool-4trp} and~\ref{fig:test-4trp}.
Figure~\ref{fig:bool-test-4trp}.\\

% The subpuzzle TEST tests whether the function value computed by the
% circuit is \emph{true} or not.  This subpuzzle has only one valid solution,
% namely that its input is blue (which means that the circuit evaluates
% to \emph{true}).
% 
% Obviously, neither of these subpuzzles, BOOL and TEST, do require
% any modification, and they are the only subpuzzles
% from~\cite{hol-hol:j:tantrix} not modified.  For completeness, we
% present them in Figures~\ref{fig:bool-4trp} and~\ref{fig:test-4trp}.
% 

% \begin{figure}[h!]
%   \centering
%   \subfigure[Output]{
%     \label{fig:test-4trp-t}
%     \quad
%     \quad
%     \includegraphics[height=1.8cm]{../Bilder-EPS-sw/test-4trp-t}
%     \quad
%     \quad
%   }       
%   \subfigure[Scheme]{
%     \label{fig:test-4trp-s}
%     \quad
%     \quad
%     \includegraphics[height=1.8cm]{../Bilder-EPS-sw/test-4trp-s}
%     \quad
%     \quad
%   }      
%   \caption{Subpuzzle TEST, see~\cite{hol-hol:j:tantrix}}
%   \label{fig:test-4trp}
% \end{figure}

The shapes of our modified subpuzzles have changed slightly, so it
might be possible that unintended interactions between two neighboring
subpuzzles do occur.  However, as noted by Holzer and
Holzer~\cite{hol-hol:j:tantrix}, the minimal horizontal distance
between two wires and/or gates is at least four, and this is still
enough to prevent any unintended interactions between our modified
subpuzzles.

\subsection{Proof of Theorem~\ref{thm:sharptrp-is-sharpp-complete}}
\label{sec:proof}

We are now ready to prove
Theorem~\ref{thm:sharptrp-is-sharpp-complete}.
Let $\sat$ denote the satisfiability problem.

\begin{lemma}
\label{lem:sat}
$\sat$ parsimoniously reduces to~$\circuitandnotsat$.
\end{lemma}

\begin{proofs}
Note that the problems $\sat$ and $\circuitsat$
(which is the same as $\circuitandnotsat$ except with OR gates allowed
as well) are equivalent under parsimonious
reductions~\cite{pap:b-1994:complexity}.  Since OR gates can be
expressed by AND and NOT gates without changing the number of
solutions, this gives a parsimonious reduction from $\sat$ to
$\circuitandnotsat$.~\end{proofs}

Now, the parsimonious reduction from $\sat$ to $\trp$ immediately
follows from Lemma~\ref{lem:sat} and the construction and the
arguments presented in Sections~\ref{sec:wire}, \ref{sec:gate},
and~\ref{sec:input-output}.  That is, by our modifications, for each
satisfying assignment to the circuit there is exactly one solution to
the Tantrix\texttrademark\ puzzle constructed.

\section{The Unique Tantrix\texttrademark\ Rotation Puzzle Problem is
DP-Complete under Randomized Reductions}
\label{sec:uniquetrp-is-dp-complete}

Valiant and Vazirani introduced \emph{randomized polynomial-time
reductions} in their work showing that $\np$ is as easy as detecting
unique solutions~\cite{val-vaz:j:np-unique}.  We will use
$\randomized$ to denote their type of reductions.  
In particular, Valiant and
Vazirani~\cite{val-vaz:j:np-unique} proved that $\usat$, the unique
version of $\sat$, is $\randomized$-complete in~$\DP$ (see also Chang,
Kadin, and
Rohatgi~\cite{cha-kad-roh:uniquesat-randomized-reductions}).

% \begin{definition}[Valiant and Vazirani~\cite{val-vaz:j:np-unique}]
%%% JR: Perhaps we can omit giving the formal definition here.
% \end{definition}

%%% JR: see email from DB of March 15, 2007:
% (1) USAT is complete in DP under randomized polynomial time 
% reduction (nach Valiant und Vazirani)
% (2) Wenn man fuer ein A in NP zeigen kann, dass SAT parsimonious 
% reducible to A ist, dann gilt insbesondere:
% USAT is parsimoniously reducible to UA.
% Wegen (1) gilt dann auch: UA ist DP-hart under randomized 
% polynomial time reduction.
% 
% Fuer Tantrix heisst das:
% SAT is parsimoniously reducible to TANTRIX (da Boolescher 
% Schaltkreis aequivalent zu einer booleschen Formel ist)
% Also gilt insbesondere auch USAT is parsimoniously reducible to 
% UTANTRIX, also ist UTANTRIX DP-hart under randomized 
% polynomial time reduction.
% Da UTANTRIX ebenfalls in DP ist, gilt somit auch UTANTRIX ist DP-
% vollstaendig.

\begin{theorem}
\label{thm:uniquetrp-is-dp-complete}
\begin{enumerate}
\item $\usat$ parsimoniously reduces to~$\utrp$.
\item $\utrp$ is $\DP$-complete under $\randomized$-reductions.
\end{enumerate}
\end{theorem}

\begin{proofs}
To prove the first part, note that by Lemma~\ref{lem:sat} and
Theorem~\ref{thm:sharptrp-is-sharpp-complete}, we obtain a
parsimonious reduction from $\sat$ to~$\trp$.  It follows that $\usat$
parsimoniously reduces to~$\utrp$.

The second part follows from the first part and Valiant and Vazirani's
above-mentioned result that $\usat$ is $\randomized$-complete
in~$\DP$, and from the obvious fact that $\utrp$ is in~$\DP$.~\end{proofs}

\bigskip

\noindent
{\bf Acknowledgments:} We thank the anonymous MCU~2007 referees for
their helpful comments.

\bibliographystyle{alpha}
%\bibliographystyle{fundam}
%\bibliography{/u/rothe/BIGBIB/joergbib}
%\bibliography{joergbib}

\bibliography{/home/rothe/BIGBIB/joergbib}

\end{document}

%% file: coordinate-system-4color.eepic
\setlength{\unitlength}{0.00053745in}
\begingroup\makeatletter\ifx\SetFigFont\undefined%
\gdef\SetFigFont#1#2#3#4#5{%
  \reset@font\fontsize{#1}{#2pt}%
  \fontfamily{#3}\fontseries{#4}\fontshape{#5}%
  \selectfont}%
\fi\endgroup%
{\renewcommand{\dashlinestretch}{30}
\begin{picture}(4094,3308)(0,-10)
\thicklines
\put(2047,1648){\blacken\ellipse{90}{90}}
\put(2047,1648){\ellipse{90}{90}}
\put(2047,2728){\blacken\ellipse{90}{90}}
\put(2047,2728){\ellipse{90}{90}}
\put(2047,568){\blacken\ellipse{90}{90}}
\put(2047,568){\ellipse{90}{90}}
\put(2992,2188){\blacken\ellipse{90}{90}}
\put(2992,2188){\ellipse{90}{90}}
\put(1102,2188){\blacken\ellipse{90}{90}}
\put(1102,2188){\ellipse{90}{90}}
\put(1102,1108){\blacken\ellipse{90}{90}}
\put(1102,1108){\ellipse{90}{90}}
\put(2992,1108){\blacken\ellipse{90}{90}}
\put(2992,1108){\ellipse{90}{90}}
\path(2362,1108)(2672,1651)(2357,2191)
	(1732,2188)(1422,1645)(1737,1105)(2362,1108)
\path(3307,1648)(3617,2191)(3302,2731)
	(2677,2728)(2367,2185)(2682,1645)(3307,1648)
\path(3312,565)(3622,1108)(3307,1648)
	(2682,1645)(2372,1102)(2687,562)(3312,565)
\path(2367,25)(2677,568)(2362,1108)
	(1737,1105)(1427,562)(1742,22)(2367,25)
\path(1412,571)(1722,1114)(1407,1654)
	(782,1651)(472,1108)(787,568)(1412,571)
\path(1412,1651)(1722,2194)(1407,2734)
	(782,2731)(472,2188)(787,1648)(1412,1651)
\path(67,523)(4072,2818)
\path(67,523)(4072,2818)
\blacken\path(3893.597,2646.616)(4072.000,2818.000)(3833.934,2750.733)(3893.597,2646.616)
\path(4027,523)(22,2818)
\path(4027,523)(22,2818)
\blacken\path(260.066,2750.733)(22.000,2818.000)(200.403,2646.616)(260.066,2750.733)
\path(2362,2188)(2672,2731)(2357,3271)
	(1732,3268)(1422,2725)(1737,2185)(2362,2188)
\put(3937,2323){\makebox(0,0)[lb]{\smash{{\SetFigFont{11}{7.2}{\rmdefault}{\mddefault}{\updefault}$x$}}}}
\put(22,2323){\makebox(0,0)[lb]{\smash{{\SetFigFont{11}{7.2}{\rmdefault}{\mddefault}{\updefault}$y$}}}}
\put(1772,2968){\makebox(0,0)[lb]{\smash{{\SetFigFont{9.5}{7.2}{\rmdefault}{\mddefault}{\updefault}$(1,1)$}}}}
\put(1772,1888){\makebox(0,0)[lb]{\smash{{\SetFigFont{9.5}{7.2}{\rmdefault}{\mddefault}{\updefault}$(0,0)$}}}}
\put(1602,288){\makebox(0,0)[lb]{\smash{{\SetFigFont{9.5}{7.2}{\rmdefault}{\mddefault}{\updefault}$(-1,-1)$}}}}
\put(2812,1828){\makebox(0,0)[lb]{\smash{{\SetFigFont{9.5}{7.2}{\rmdefault}{\mddefault}{\updefault}$(1,0)$}}}}
\put(832,1828){\makebox(0,0)[lb]{\smash{{\SetFigFont{9.5}{7.2}{\rmdefault}{\mddefault}{\updefault}$(0,1)$}}}}
\put(777,728){\makebox(0,0)[lb]{\smash{{\SetFigFont{9.5}{7.2}{\rmdefault}{\mddefault}{\updefault}$(-1,0)$}}}}
\put(2592,728){\makebox(0,0)[lb]{\smash{{\SetFigFont{9.5}{7.2}{\rmdefault}{\mddefault}{\updefault}$(0,-1)$}}}}
\end{picture}
}